\documentclass[pdftex]{iopart}
\usepackage{graphicx}

\begin{document}

\title[Measuring EAS with Cherenkov light detectors]{Measuring extensive air showers with Cherenkov light detectors of the Yakutsk array: The energy spectrum of cosmic rays}

\author{A.A.~Ivanov, S.P.~Knurenko and I.Ye.~Sleptsov}

\address{Shafer Institute for Cosmophysical Research and Aeronomy, Yakutsk 677980, Russia}
\ead{ivanov@ikfia.ysn.ru}

\begin{abstract}
The energy spectrum of cosmic rays in the range $E\sim 10^{15}$ eV to $6\times 10^{19}$ eV has been studied using the air Cherenkov light detectors of the Yakutsk array. The total flux of photons produced by relativistic electrons (including positrons as well, hereafter) of extensive air showers in the atmosphere is used as the energy estimator of the primary particle initiating a shower. The resultant differential flux of cosmic rays exhibits, in accordance with previous measurements, a knee and ankle features at energies $3\times10^{15}$ and $\sim10^{19}$ eV, respectively. A comparison of observational data with simulations is made in the knee and ankle regions in order to choose the models of galactic and extragalactic components of cosmic rays which describe better the energy spectrum measured.
\end{abstract}

\pacs{96.50.sd, 98.70.Sa}
\submitto{\NJP}
\maketitle

%%%%%%%%%%%%%%%%%%%%%%%%%%%%%%%%%%%%%%%%%%%%%%%%%%%%%%%%%%%%%%%%%%%%%

\section{Introduction}
The Cherenkov light emitted in the atmosphere by relativistic electrons of extensive air showers (EASs) of cosmic rays (CRs) carries important information about the shower development and the primary CR particles. Well-known application of the air Cherenkov light technique is the $\gamma$-ray astronomy. But in this article we aim at another application of the technique - namely to measure Cherenkov photons from EAS initiated by CR of energy above $10^{15}$ eV. These giant showers supply plenty of light so that one can detect it with unarmed photomultiplier tubes (PMT) triggered within pulse duration.

Since the first observation by Galbraith and Jelley~\cite{Jelley} and a systematic measurement of air Cherenkov light properties in the Pamir experiment~\cite{Chudakov}, a number of EAS arrays have been equipped with Cherenkov light detectors. It seems that the most durable and plentiful in Cherenkov light data is the Yakutsk array experiment~\cite{Mono,CERN}. The total flux of light is used to estimate the primary energy in a model independent manner and the radial distribution of the light intensity at ground level is used to infer the position of shower maximum, $X_{max}$, in the atmosphere~\cite{Mono}.

In this article we focus on the experimental data obtained in Yakutsk with Cherenkov light detectors aiming at the CR energy spectrum. The paper is structured as follows. In Section~\ref{Sctn:2} we outline the general characteristics of the Yakutsk array experiment, while in Sections~\ref{Sctn:3} and~\ref{Sctn:4} particular properties of the Cherenkov light detectors are given: detector design and calibration of the signal. In Section~\ref{Sctn:5} the measurement and monitoring of the atmospheric extinction of light is described. The measured lateral distribution of the Cherenkov light intensity at the observation level is given in Section~\ref{Sctn:6}. The method used to estimate the energy of the particle initiating the EAS is described in Section~\ref{Sctn:7}. Resultant energy spectrum of CRs is discussed in Section~\ref{Sctn:8}. Our conclusions are set out in Section~\ref{Sctn:9}. In two Appendices additional material is given essential for the subjects considered.

%%%%%%%%%%%%%%%%%%%%%%%%%%%%%%%%%%%%%%%%%%%%%%%%%%%%%%%%%%%%%%%%%%%%%

\section{The Yakutsk array}\label{Sctn:2}
The Yakutsk array is located in Oktyomtsy near Yakutsk, Russia ($61.7^0N,129.4^0E$), 100 m above sea level (1020 g/cm$^2$). At present it consists of 58 ground-based and 6 underground scintillation detector stations to measure charged particles (electrons and muons) and 48 detectors - PMTs in shuttered housing to observe the atmospheric Cherenkov light. During more than 30 years of lifetime the Yakutsk array has been re-configured several times, the total area covered by detectors was maximal about 1990 ($S_{eff}\sim17$ km$^2$), now it is $S_{eff}\sim10$ km$^2$. In the central part of the array there is a denser domain with 100-250 m detector spacing. During the whole observation period approximately $10^6$ showers of the primary energy above $10^{15}$ eV are detected; the three highest energy events selected with axes within the array area and zenith angle $\theta\leq60^0$ have an energy $E>10^{20}$ eV.

\begin{figure}
\center{\includegraphics[width=0.6\columnwidth]{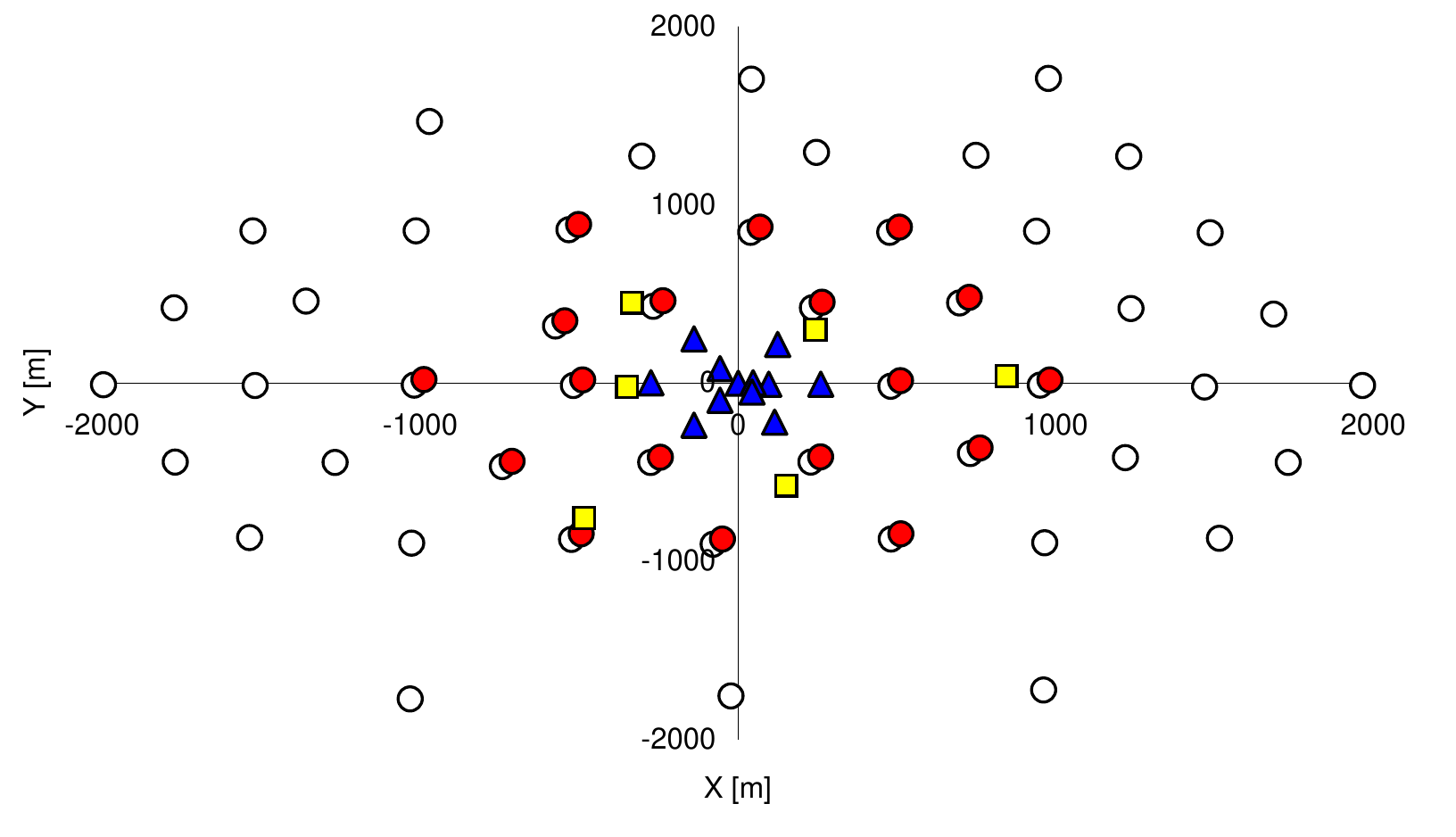}}
\caption{The detector arrangement of the Yakutsk array. Charged particle detectors (open circles), Cherenkov light detectors of the $C_1$ subset (filled circles) and the $C_2$ subset (filled triangles), and the muon detectors (squares) are shown.}
\label{Fig:Map}
\end{figure}

The actual detector arrangement of the array is shown in Figure~\ref{Fig:Map}. Charged particle detectors of 2 m$^2$ area are built in stations in couples; the Cherenkov light detectors - PMTs of 176 cm$^{2}$ and $3\times176$ cm$^{2}$ acceptance area, forms the medium, $C_1$ ($\sim500$ m spacing), and the autonomous, $C_2$ (50 to 200 m spacing), subsets. The latter was added in 1995 with the aim to study air showers in the energy range $10^{15}-10^{17}$ eV via the Cherenkov light measurements~\cite{Autonom}.

\begin{figure}
\center{\includegraphics[width=0.6\columnwidth]{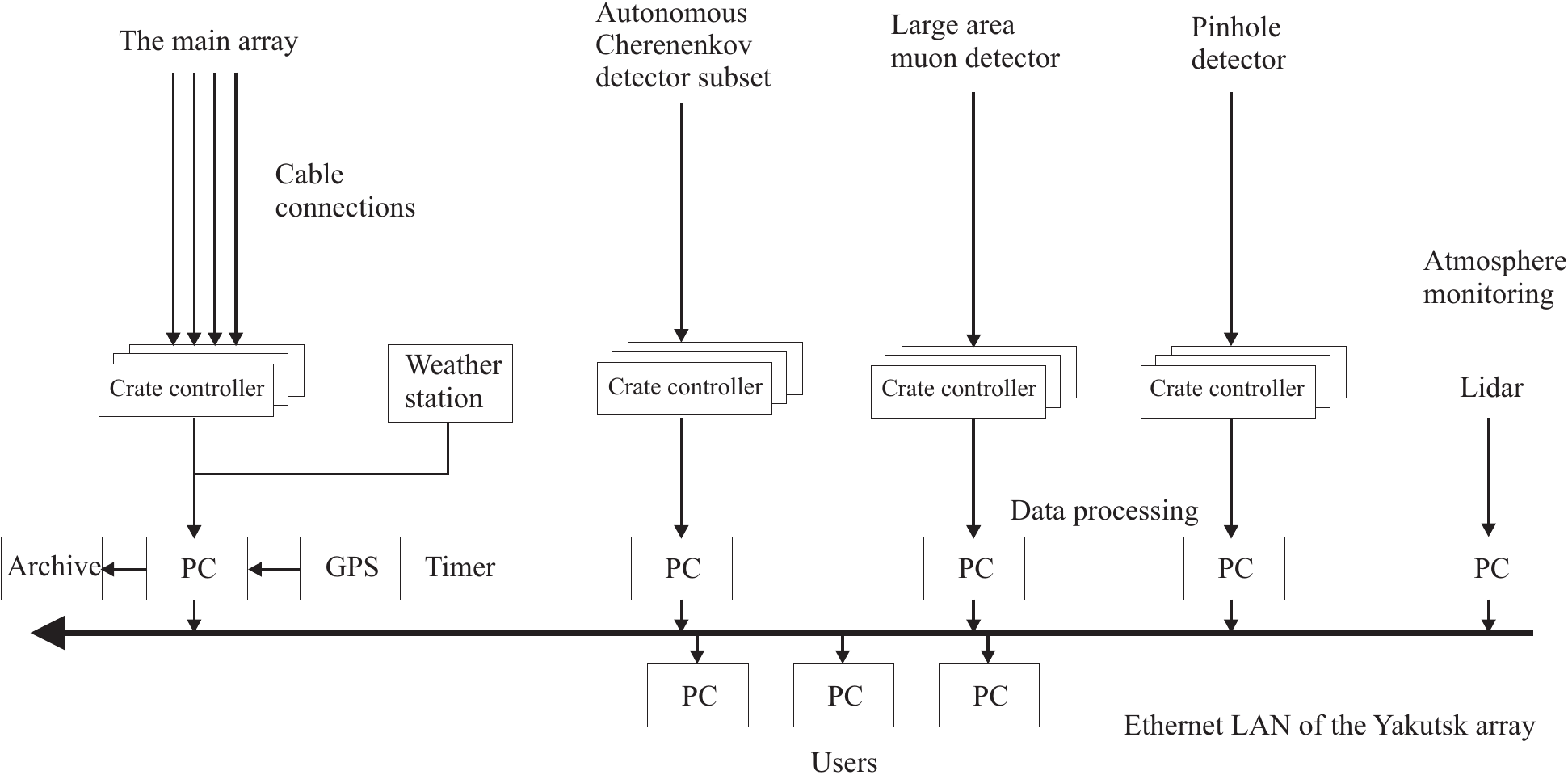}}
\caption{Local area network of the Yakutsk array.}
\label{Fig:LAN}
\end{figure}

All detectors/controllers and data processing units of the array are connected into the data handling network shown in figure~\ref{Fig:LAN}.

%%%%%%%%%%%%%%%%%%%%%%%%%%%%%%%%%%%%%%%%%%%%%%%%%%%%%%%%%%%%%%%%%%%%%%%%%%%%%%%%%%%%%%%%%%%%%%%%%%%%%%%%%%

\section{Air Cherenkov light detector design}\label{Sctn:3}
\begin{figure}
\center{\includegraphics[width=0.6\columnwidth]{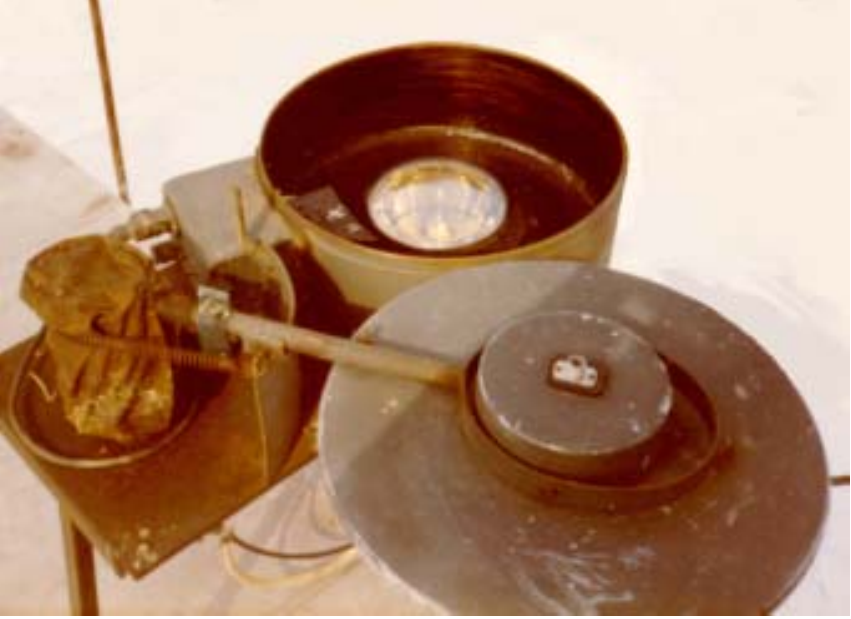}}
\caption{The air Cherenkov light detector.}
\label{Fig:CherDet}
\end{figure}

Charged particle detectors of the array were described in \cite{Mono}. Here we will detail a Cherenkov light detector unit. It consists of a vertically mounted PMT (FEU-49B, 15 cm diameter) with amplifier in a metal container blackened inside \cite{Sleptsov}. An upper hole provides $\theta\leq55^0$ aperture (figure~\ref{Fig:CherDet}). To protect the photocathode from sunlight the motorized light-proof lid is set. At night, all lids of the array can be commanded remotely to open. PMTs and amplifiers are powered around-the-clock to guarantee a stability of performance. When the lid is open, a fan blows with warm air to keep snow and dust out of the photocathode surface.

\begin{figure}
\center{\includegraphics[width=0.6\columnwidth]{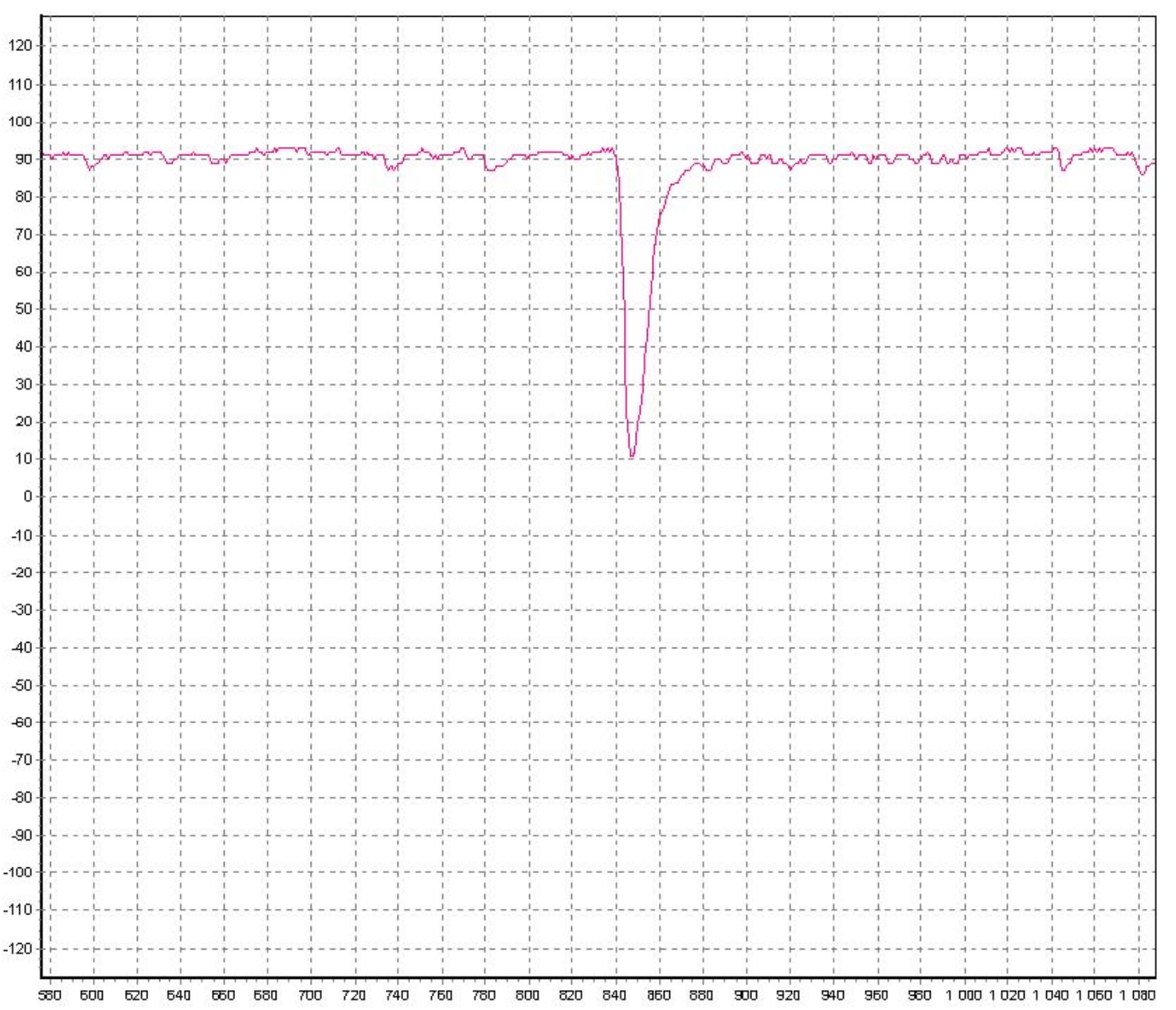}}
\caption{Time (horizontal axis, ns) dependence of the Cherenkov light signal (arbitrary units, vertical axis).}
\label{Fig:PulseShape}
\end{figure}

\begin{figure}
\center{\includegraphics[width=0.6\columnwidth]{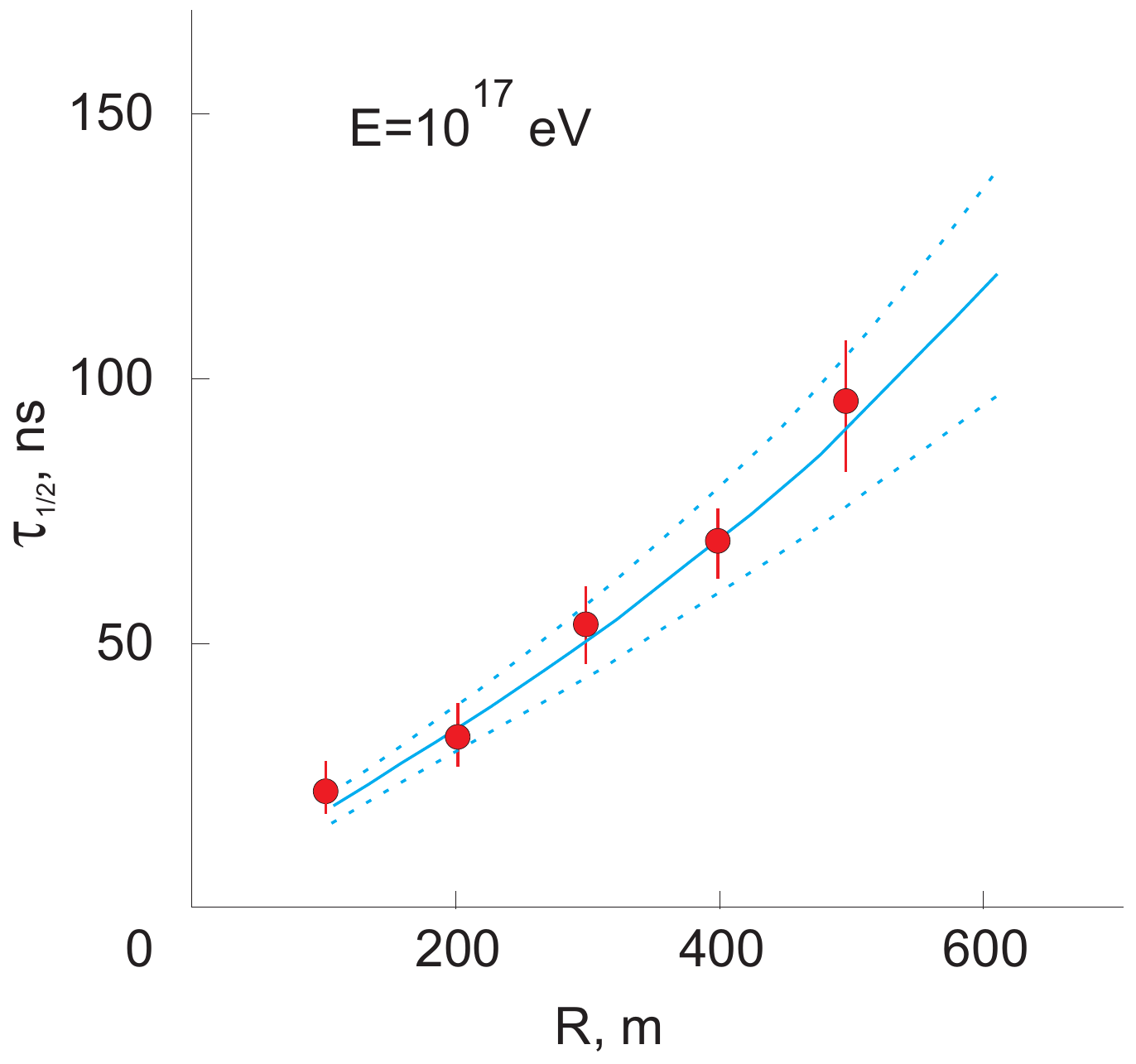}}
\caption{The half width of the Cherenkov light pulse as a function of the shower core distance. Experiment: points~\cite{Mono}; theory: solid line with $1\sigma$ errors (dashed lines)~\cite{Hara}.}
\label{Fig:HalfWidth}
\end{figure}

There is a variant of detector with three PMTs in a housing which can operate independently or in summation of signal in order to increase acceptance area at the shower periphery. In addition, dedicated detectors were used to measure the shape and width of the Cherenkov signal from the shower. As an example, the pulse shape of the Cherenkov signal at the shower periphery is shown in figure~\ref{Fig:PulseShape}, while the halfwidth of the signal as a function of radial distance is given in figure~\ref{Fig:HalfWidth}.

\begin{figure}
\center{\includegraphics[width=0.6\columnwidth]{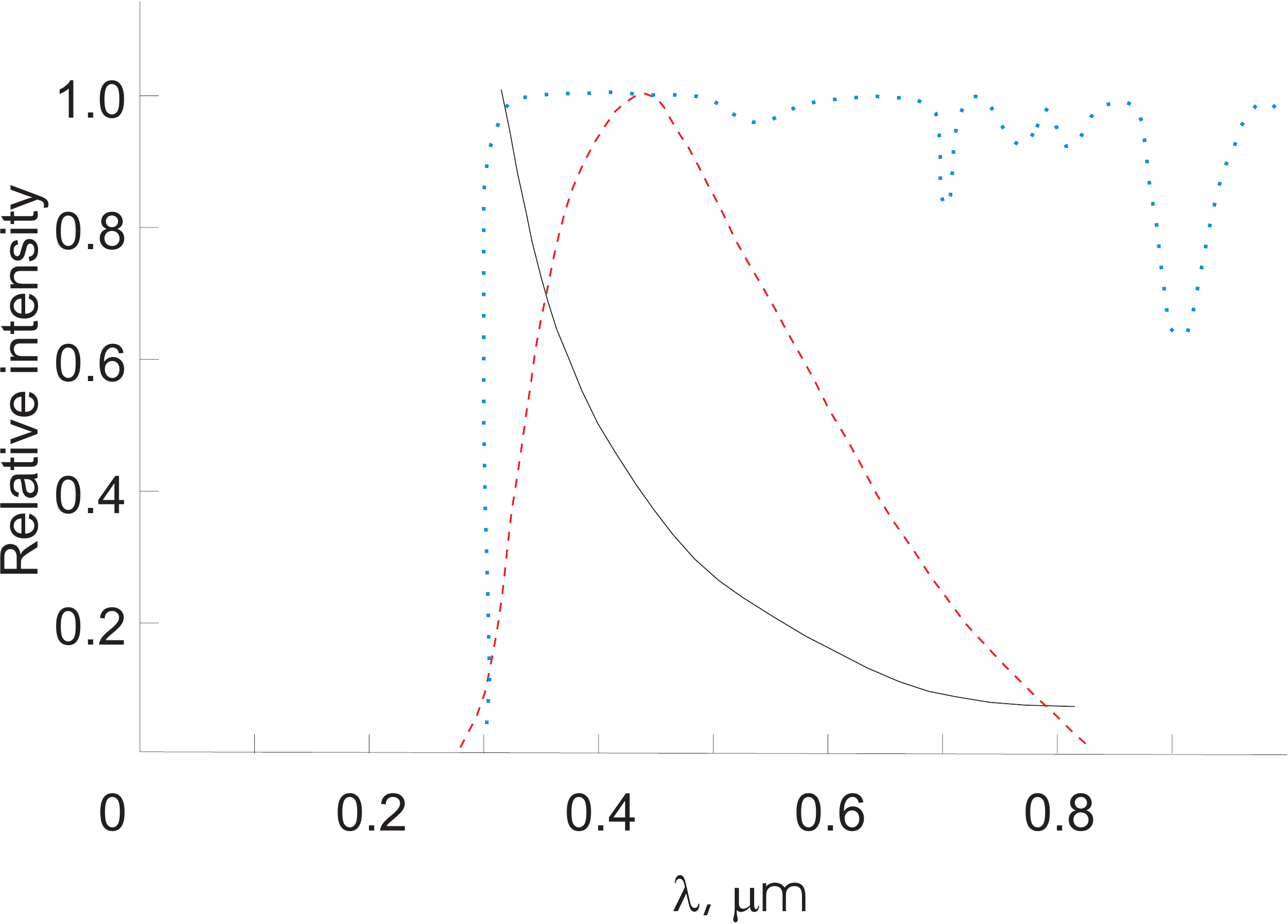}}
\caption{The spectral sensitivity of the PMT (dashed line). Air transmission of a cloudless atmosphere (dotted line) and Cherenkov emission spectrum (full line).}
\label{Fig:Sens}
\end{figure}

The spectral sensitivity of the PMT used is shown by the dashed line in figure~\ref{Fig:Sens} together with air Cherenkov light spectrum (solid line) and the atmospheric transmission curve.

Two kind of triggers were used to select the showers from the background at the Yakutsk array: produced by the scintillators and by Cherenkov light detectors. In the first case a coincidence signal (if the particle density, $\rho$, is greater than $0.5$ m$^{-2}$ in two scintillators of each station within 2~$\mu$s) passes on to the central controller. Trigger-500 is then produced in the case of a coincident signal (in 40~$\mu$s) from three or more stations with $\sim500$ m spacing ($\geq$~sixfold coincidence). Similarly, trigger-1000 is produced by $\sim 1$ km spacing stations. After 1992 when 18 new stations were added, the array area is increased from 2.5 km$^2$ to 7.2 km$^2$ where trigger-500 operates. That is why we can deal with EAS in the energy range from $3\times10^{16}$ eV to $3\times10^{19}$ eV using the same trigger.

The $C_1$ subset of Cherenkov light detectors has no hardware trigger of its own. Instead, the scintillator trigger controls the detection of Cherenkov signal in this subarray. The amplitude of PMT signal is recorded within 5 $\mu$s after the coincident signal in two scintillators of the host station. In order to exclude detection of an accidental Cherenkov signal when there is no trigger-500/1000,
a double trigger condition is in use in data analysis:
\begin{itemize}
 \item three or more scintillator stations have detected the charged particles ($\rho>0.5$ m$^{-2}$);
 \item the Cherenkov light intensities in three or more PMTs are greater than $2.4\times 10^{5}/4.5\times 10^{5}$ m$^{-2}$ depending on the acceptance of PMT.
\end{itemize}
The thresholds have been chosen to keep the signal/noise ratio greater than 3.

The second case is $C_2$ subset which has an independent Cherenkov light trigger formed by three or more PMTs having detected light intensities above a given threshold within 10 $\mu$s. The signal integration time of the individual PMT is 0.5 $\mu$s.

%%%%%%%%%%%%%%%%%%%%%%%%%%%%%%%%%%%%%%%%%%%%%%%%%%%%%%%%%%%%%%%%%%%%%%%%%%%%%%%%%%%%%%%%%%%%%%%%%%%%%%%%%%

\section{Detector calibration}\label{Sctn:4}
\begin{figure}
\center{\includegraphics[width=0.6\columnwidth]{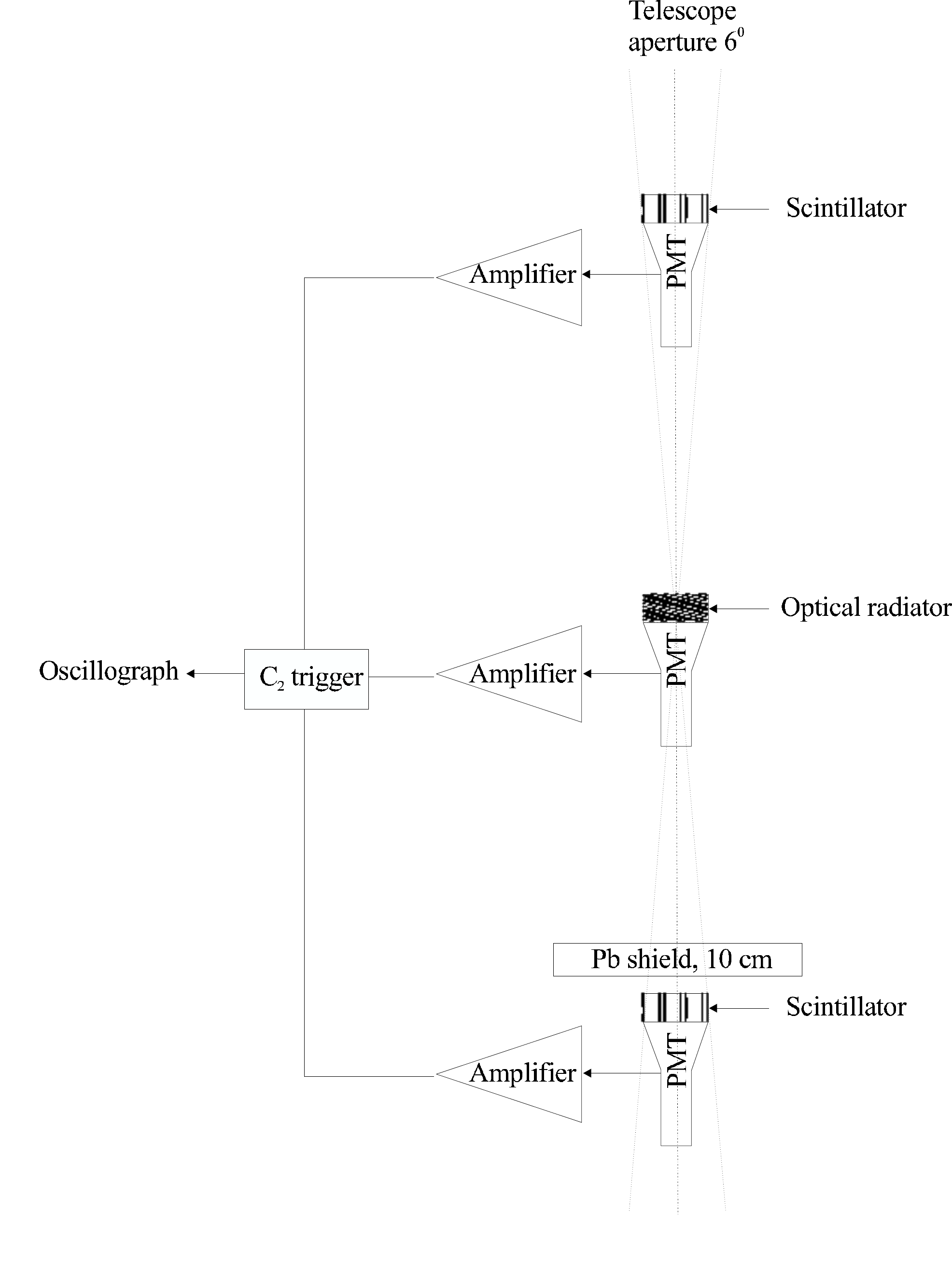}}
\caption{A setup to calibrate the air Cherenkov light detector.}
\label{Fig:Calibration}
\end{figure}

In order to find the absolute response of the array to Cherenkov light we have considered the sensitivity of all the PMTs in the detectors and have measured the conversion factors from light intensity into ADC value. The Cherenkov light emitted by relativistic muons in a transparent medium with known refraction index (optical radiator) was used in the detector calibration, as was suggested in~\cite{Nesterova}. In this paper we are keeping along the description given in~\cite{Sleptsov, Glushkov}. Distilled water and plexiglas are used to measure the output signal in a geometry shown in figure~\ref{Fig:Calibration}. A telescope composed of two PMTs above and below the radiator, spaced 1 m from the detector, bottom one under the lead shield (10 cm), selects high-energy muons from background CRs. On the photocathode of the detector PMT is placed (above a thin layer of glycerin) a plexiglas disk of 5.5 cm thickness and 15 cm diameter with polished side surface. As another version of radiator, distilled water was used without any pan layer. The Cherenkov radiation angle in water is $35^0$, while it is greater in plexiglas ($48^0$) and in glycerin ($47^0$).

The photon number in the wavelength interval ($\lambda_1=3000 {\AA},\lambda_2=8000 {\AA}$) emitted in the radiator ($n=1.49, l=5.5$ cm) is given by the Frank and Tamm formula~\cite{Tamm}:
\begin{equation}
\frac{dN_\gamma}{dE_\mu}=2\pi\alpha l(\frac{1}{\lambda_1}-\frac{1}{\lambda_2})(1-\frac{E_\mu^2}{n^2(E_\mu^2-m_\mu^2)}),
\label{Eq:Tamm}
\end{equation}
where $\alpha=1/137$; $n$ is refraction coefficient; $m_\mu,E_\mu$ are muon mass and energy. Integrating it above the threshold energy ($E_\mu>0.25$ GeV) with the muon spectrum measured at sea level we have $N_\gamma=2850$.
Then the ratio of the air Cherenkov light intensity $Q(R)$ at distance $R$ from the shower core to $N_\gamma$ is
$$
\frac{Q(R)S}{N_\gamma}=\frac{A_s \overline{K_r}}{A_r(1-R_s)},
$$
where $A_s,A_r$ are the detector signal amplitudes from the shower and radiator; S is photocathode area; $R_s\leq0.04$ is the reflectivity factor of air-glass junction.
$$
\overline{K_r}=\frac{\int_{\lambda_1}^{\lambda_2}N_\gamma(\lambda)S(\lambda)
K_r(\lambda)d\lambda}
{\int_{\lambda_1}^{\lambda_2}N_\gamma(\lambda)S(\lambda)d\lambda}=0.93,
$$
where $K_r(\lambda)$ is a radiator transparency of effective thickness 4 cm (taking into account a radiation cone). The spectral characteristics of PMT, plexiglas and glycerin are measured with the spectrophotometer. Similar values were derived in the case of water radiator~\cite{Glushkov}.

Optical radiators used in the laboratory for absolute calibration of the detector produce photon number insufficient for the routine ratio calibration of a multitude of array detectors in the field. Instead, the scintillator disk (D=15 cm, h=5 cm) is in use as the light source in this case. Intercalibration of different radiators has been performed using the experimental setup (figure~\ref{Fig:Calibration}).

The total calibration error of 21\% consists of uncertainties in:
 \begin{itemize}
 \item number of photons at photocathode (10\%);
 \item measurement of signal amplitude from radiator (9\%);
 \item conversion factor from radiator to scintillator (7\%);
 \item signal variance throughout the photocathode area and acceptance angle (15\%).
 \end{itemize}

%%%%%%%%%%%%%%%%%%%%%%%%%%%%%%%%%%%%%%%%%%%%%%%%%%%%%%%%%%%%%%%%%%%%%%%%%%%%%%%%%%%%%%%%%%%%%%%%%%%%%%%%%%

\section{Atmospheric extinction of light}\label{Sctn:5}
Cherenkov light undergo extinction in the atmosphere because of absorption on molecules by Rayleigh scattering and Mie scattering by aerosols and absorption. During the observation periods in winter (average temperature $-40^0$~C) Cherenkov light absorption in the atmosphere is negligible and only molecular and aerosol scattering of photons are taken into account.

Molecular scattering is almost constant, while aerosol concentration in the boundary layer above the surrounding terrain is of diurnal and seasonal variability. We are using the event rate of the showers in the energy range $10^{15}-10^{16}$ eV detected with a PMT subset in hour and 15 min intervals in order to monitor the atmospheric transparency for the light generated in EAS~\cite{Mono}. The method is based on the Cherenkov light flux proportional to the primary energy:
\begin{equation}
\tau_i=\tau_0(\frac{N_i(>Q_{thr})}{N_0(>Q_{thr})})^{1/\kappa},
\label{Eq:Sokurov}
\end{equation}
where $N(>Q_{thr})$ is integral number of events with the light intensity above the threshold detected in $i$-th and basic periods; $\tau$ is atmospheric extinction coefficient; $\kappa$ is the energy spectrum index.

Resultant extinction coefficient is a product of variable $\tau_a$ due to current aerosol concentration and basic coefficient $\tau_0$ caused by the molecular scattering and minimal aerosol extinction. Rayleigh scattering parameters evaluated for the Yakutsk array conditions give $\tau_R=0.9$~\cite{Sleptsov} and the corresponding attenuation length $\lambda_R=300$ km, averaged over wavelength range under consideration.

The integral spectrum index of the Cherenkov light flux around the knee is~\cite{Sleptsov}:
\begin{eqnarray}
\kappa=1.5\pm0.03,\;E_0<3\times10^{15},\nonumber\\
\kappa=2.1\pm0.04,\;E_0>3\times10^{15}.\nonumber
\end{eqnarray}

\begin{figure}
\center{\includegraphics[width=0.6\columnwidth]{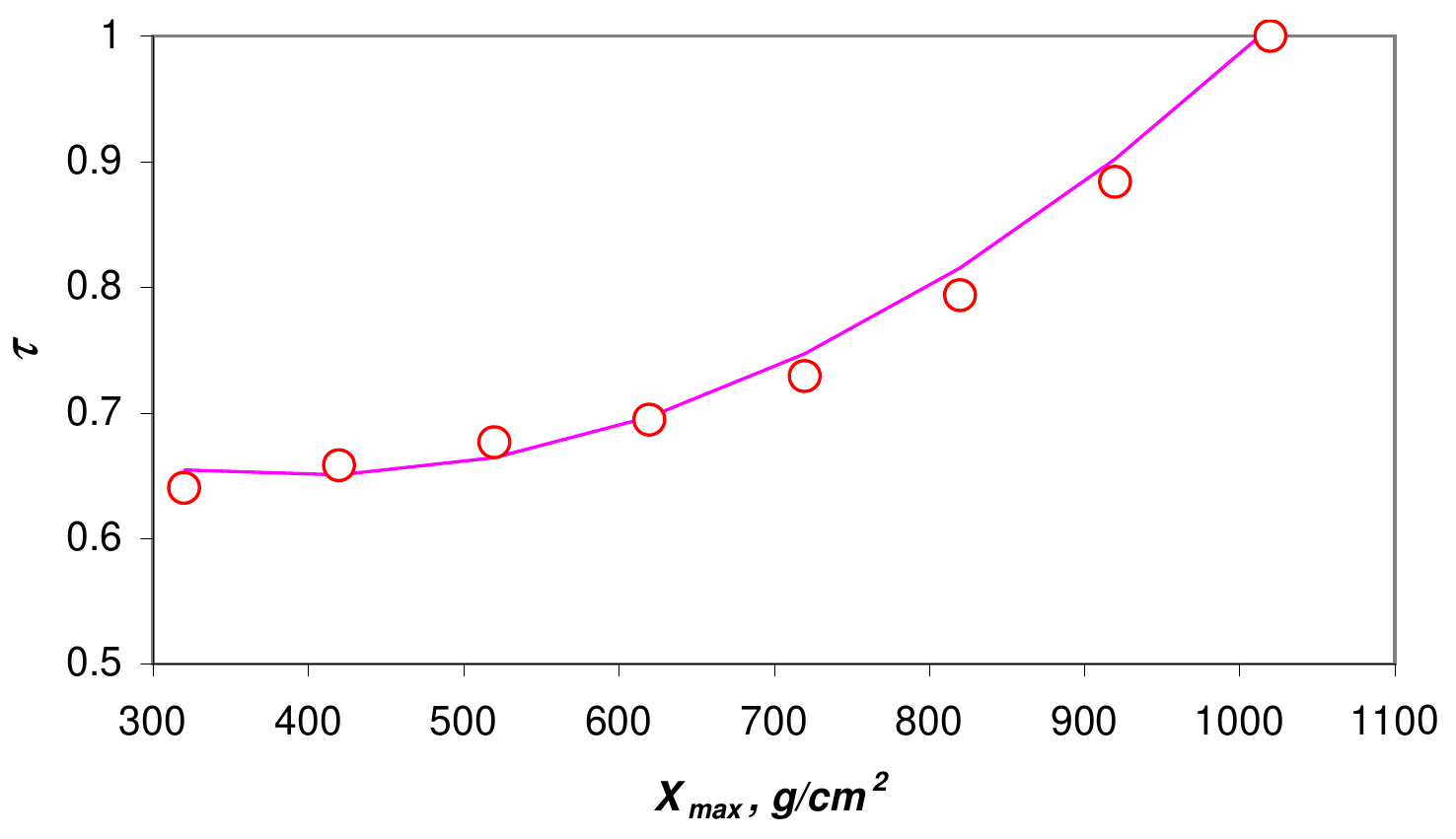}}
\caption{Atmospheric extinction of the Cherenkov light as a function of $X_{max}$. $\theta=20^0$ (circles). Approximation~\ref{Eq:Transparency} is shown by the solid line.}
\label{Fig:Transparency}
\end{figure}

Evaluating the coefficient $\tau_0$ as a function of $X_{max}$ in the shining point approximation~\cite{LaJolla} by assuming the attenuation length due to aerosol scattering constant above 1.5 km height and decreasing below~\cite{Pampa}, one can derive the average extinction coefficient shown in figure~\ref{Fig:Transparency}.
The approximation is
\begin{equation}
\tau=(0.79\pm0.02)-\frac{X_{max}}{1400\pm30}+\frac{X_{max}^2}{(1.11\pm0.02)\times10^6},
\label{Eq:Transparency}
\end{equation}
where $X_{max}$ is in g/cm$^2$.

Finally, we have found
$$
\overline{\tau}=0.765\pm0.015
$$
at $\overline{\theta}=20^0,\overline{E}=7\times 10^{16}$ eV averaged over all observation periods, and inserting spectrum index in equation~(\ref{Eq:Sokurov}).

%%%%%%%%%%%%%%%%%%%%%%%%%%%%%%%%%%%%%%%%%%%%%%%%%%%%%%%%%%%%%%%%%%%%%%%%%%%%%%%%%%%%%%%%%%%%%%%%%%%%%%%%%%

\section{Lateral distribution of air Cherenkov light}\label{Sctn:6}
During $\sim15000$ hours of observation $\sim60000$ showers of energy above $6\times 10^{16}$ eV were detected by the medium $C_1$ subset. The autonomous $C_2$ array data consist of $\sim 200000$ showers with $E>1.2\times 10^{15}$ eV detected during $\sim 3200$ hours of observation. The Cherenkov trigger condition only is used to select these showers.

Relative dispersion of the light intensity observed with different detectors at $R\sim500$ m from the shower core has been evaluated selecting showers with 3 detectors hit in the interval $400<R<500$ m not far from each other ($\leq75$) m. Intensity dispersion contains, in addition to inherent fluctuations instrumental errors, core location error, intercalibration error, etc. Supposing chance variation of the signal around an average $Q(R)\propto R^{-2.35}$ \cite{DensFluct}, we have found $\delta Q/Q=0.25\pm0.13$. In order to minimize the $Q(R)$ uncertainty due to EAS axis location error, one has to select showers with axes within $R<R_{opt}$ m, ($80<R_{opt}<250$, if $E_0>2\times10^{17}$ eV; $80<R_{opt}<400$, if $E_0>2\times10^{18}$ eV) where stations are spaced closely.

\begin{figure}
\center{\includegraphics[width=0.6\columnwidth]{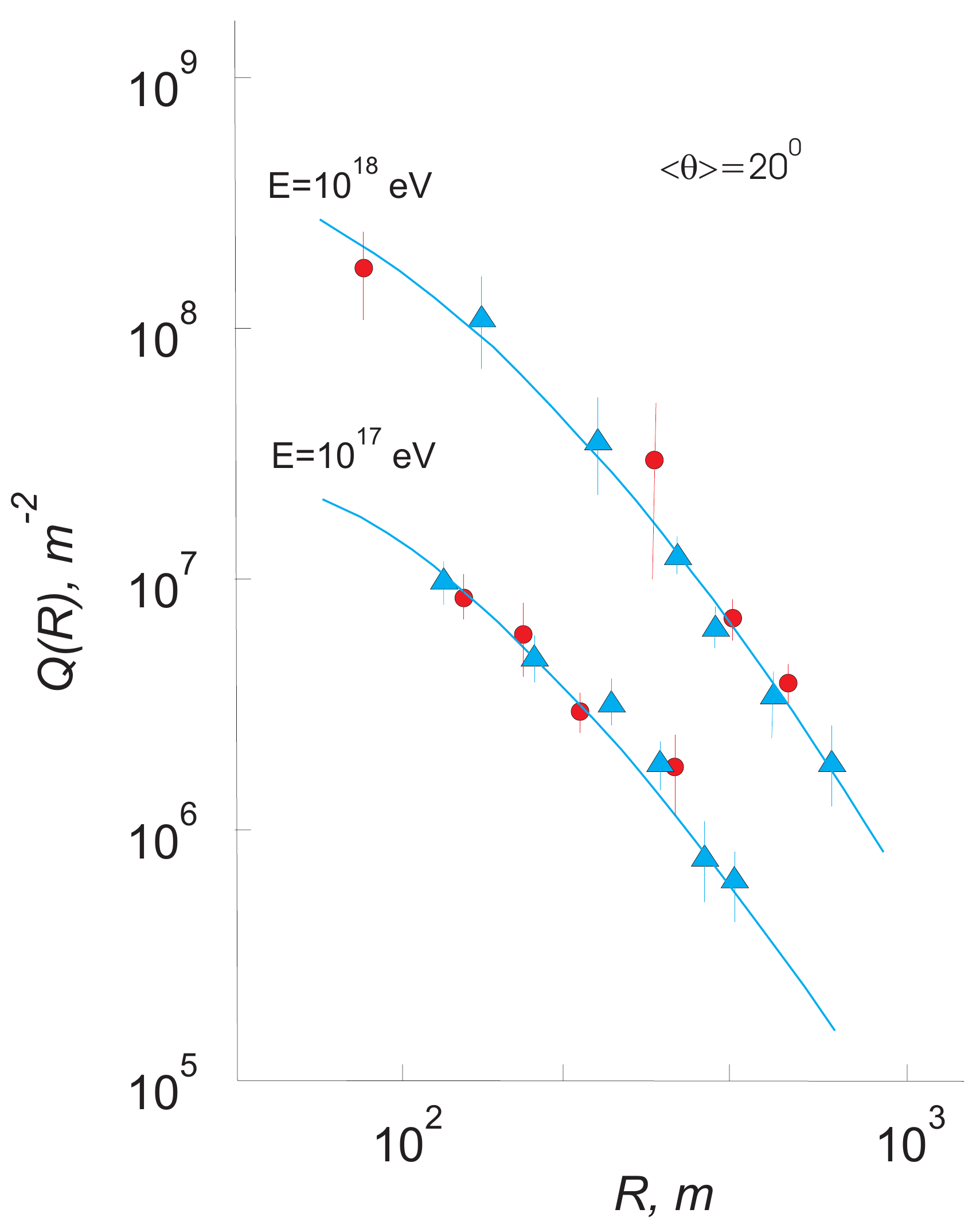}}
\caption{Comparison of lateral distribution of air Cherenkov light. Triangles - this measurement; circles from ref~\cite{HPCher}.}
\label{Fig:LDFcomparison}
\end{figure}

In the highest energy domain measurements of Cherenkov light exist, carried out by the Haverah Park group~\cite{HPCher}. In figure~\ref{Fig:LDFcomparison} our results are given in comparison with~\cite{HPCher}. The data are normalized at 200 m core distance and exhibit consistency in the lateral distribution function (LDF) shape. The solid lines are an approximation to our data given in~\cite{Glushkov}.

\begin{figure}
\center{\includegraphics[width=0.6\columnwidth]{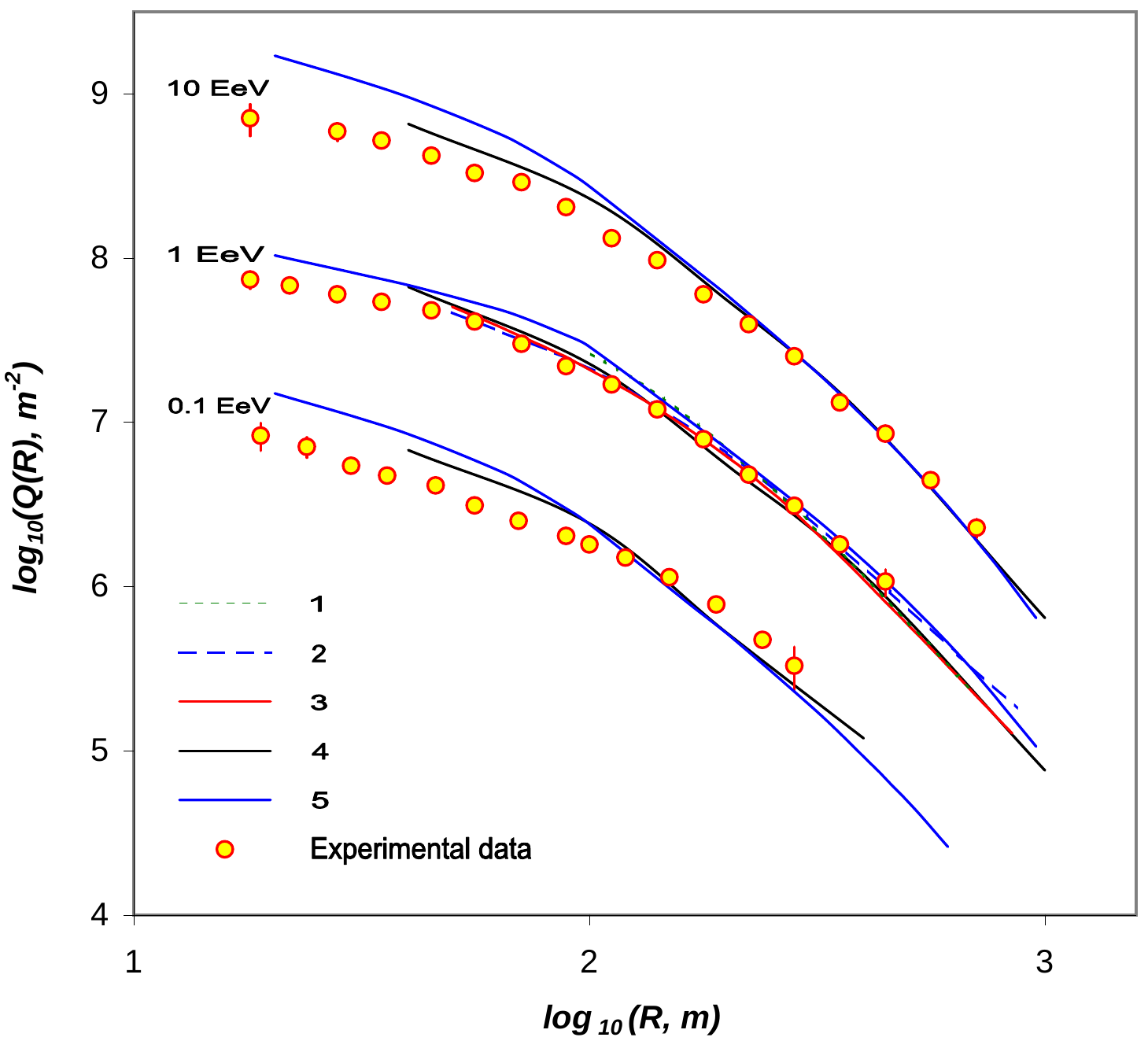}}
\caption{Results of model calculation and the Yakutsk array data (circles). The lines illustrate results of 1: Hara et al.~\cite{Hara}, 2: Dyakonov~\cite{Dyakonov}, 3: Ivanenko et al.~\cite{Ivanenko}, 4: Lagutin et al.~\cite{LagutinLDF}, and 5: Dedenko et al.~\cite{Ddnk}. The primary particle energy is indicated at the left of data/lines.}
\label{Fig:ModeLDF}
\end{figure}

Model simulations give a variety of lateral distribution functions, some of them are close to our data, as illustrated in figure~\ref{Fig:ModeLDF}. The lateral distribution depends on the attenuation lengths, multiplicity of secondaries in the interactions, primary mass composition etc., which can be parameterized with $X_{max}$. Another considerable influence is the angular distribution of electrons in the shower. The combination of these factors results in different LDFs. The results of Ivanenko et al. and Dyakonov are compatible with our measurements.

\begin{figure}
\center{\includegraphics[width=0.6\columnwidth]{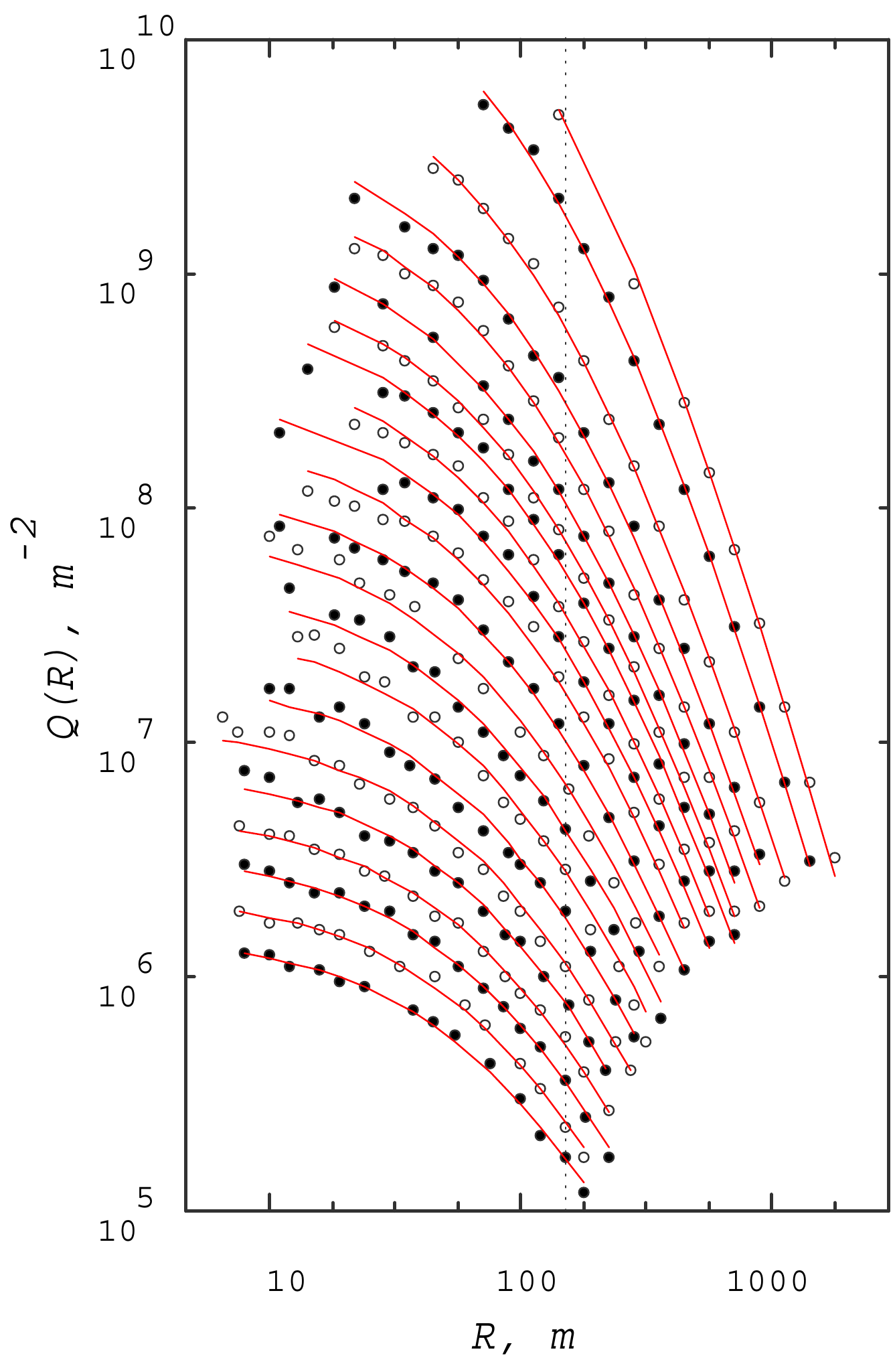}}
\caption{Air Cherenkov light radial distribution. Our experimental data are given by the black and white points alternately, not to confuse the adjacent data. The curves are approximations according to Equation~(\ref{Eq.ldf}) with Q$_{150}$ fitted to the particular data. An axis distance $R=150$ m is indicated by the dotted line.}
\label{Fig:LDF}
\end{figure}

Our cumulative results on LDF measurements (zenith angle $\theta<30^0$) are given in figure~\ref{Fig:LDF}. The data of both subsets $C_1,C_2$ are parameterized by the intensity at 150 m from the shower core, Q$_{150}$, the only core distance really present in the shifting range of measurements when the primary energy is rising from $E_0\sim 10^{15}$ to 10$^{19}$ eV. The data are consistent with the previous results of the Yakutsk array concerning the Cherenkov light and can be described by the suitable EAS model simulation ~\cite{Dyakonov,Ivanenko}.

No abrupt change of LDF parameters is seen, so we choose a rather smooth approximation curve to fit the experimental data in the whole energy range~\cite{CERN}:
\begin{equation}
Q(R)=Q_{150}\frac{(R_1+150)(R_2+R)^{1-b}}{(R_1+R)(R_2+150)^{1-b}},
\label{Eq.ldf}
\end{equation}
where $R_1=60$ m; $R_2=200$ m; $b=(1.14\pm0.06)+(0.30\pm0.02)\times \lg Q_{150}$.

%%%%%%%%%%%%%%%%%%%%%%%%%%%%%%%%%%%%%%%%%%%%%%%%%%%%%%%%%%%%%%%%%%%%%%%%%%%%%%%%%%%%%%%%%%%%%%%%%%%%%%%%%%

\section{Energy estimation}\label{Sctn:7}
The total flux of Cherenkov light emitted, $Q_{tot}$, is our main estimator of the primary particle energy. In order to derive the relation between $Q_{tot}$ and ionization loss of the shower electrons in the atmosphere we have used formula~(\ref{Eq:Tamm}) for the number of photons induced, $N_\gamma$, in ratio to that from optical radiator, $N_r$:
$$
\frac{dN_\gamma}{dh}=\frac{dN_r}{dh}\frac{1}{1-1/n_r^2}(1-(\frac{1}{vn})^2),
$$
where $v$ is electron velocity; $n,n_r$ are the refraction indexes in air and radiator, respectively; $c=1$.

Emission is possible when $vn>1$ or $E^2>E_{thr}^2=m^2/(1-1/n^2)$. The number of photons induced in the depth interval $dx$ is
\begin{equation}
\frac{dN_\gamma}{dx}=\zeta(1-\frac{E_{thr}^2}{E^2}),
\label{Eq:Dimin}\end{equation}
where $\zeta=2\frac{dN_r}{dx}\frac{\rho_r}{\rho_0}\frac{n_0-1}{1-1/n_r^2}$; $\rho_r$ is radiator density; $\rho_0,n_0$ are air density and refraction at the observation level.

The total light flux is given using the electron differential energy spectrum in the shower at the depth $x$:
\begin{equation}
Q_{tot}(x_0)=\int_0^{x_0}dx\tau(x_0-x)\int_{E_{thr}}^{E_0}dE\zeta(\frac{E_{thr}^2}{E^2}-1)\frac{dN(x,E,E_0)}{dE},
\label{Eq:Qtot}
\end{equation}
where $\tau(x_0-x)$ is the extinction coefficient of light along the path $x_0-x$.

The number of electrons in a shower is approximately $N(x,E,E_0)\simeq N(x,E_0)\chi(E)$, where $N(x,E_0)$ is the number of electrons with $E>0$; $\chi(E)$ is the universal electron spectrum at the shower maximum. Then
\begin{equation}
Q_{tot}=2\zeta E_{thr}^2\int_0^{x_0}dx\tau(x_0-x)N(x,E_0)\int_{E_{thr}}^{E_0}\chi(E)\frac{dE}{E^3}
\label{eq:QtotEi}\end{equation}
leads to the relation with the ionization loss of electrons in the atmosphere $E_i=\varepsilon_0\int_0^{x_0}N(x,E_0)dx/t_0$:

\begin{equation}
Q_{tot}^{\tau=1}\approx2\zeta t_0 (0.275-0.283\frac{E_{thr}(X_{max})}{\varepsilon_0})\frac{E_i}{\varepsilon_0},
\label{eq:QtotXmax0}
\end{equation}
if we assume the extinction $\tau=1$~\cite{JETP07}.

In the real case of the Yakutsk array conditions $\bar{T}=-30^0$ C, $P=754$ Torr and the extinction of light described in Section 5, it was found that the relation is~\cite{JETP07}:
\begin{equation}
E_i/Q_{tot}\approx(3.01\pm0.36)\times10^4(1-X_{max}/(1700\pm270))
\label{Eq:Ei2Qtot}
\end{equation}
for $X_{max}\in(500,1000)$ g/cm$^2$; the accuracy of the relation is $\sim 5\%$. It demonstrates the advantage of the air Cherenkov light measurement technique: the relation is determined by $X_{max}$ and extinction of light only; interaction model dependence is parameterized by means of $X_{max}$.

In \ref{app-a} we analyze the shower parameters governing the energy fractions transferred to EAS components.

%%%%%%%%%%%%%%%%%%%%%%%%%%%%%%%%%%%%%%%%%%%%%%%%%%%%%%%%%%%%%%%%%%%%%%%%%%%%%%%%%%%%%%%%%%%%%%%%%%%%%%%%%%

\subsection{Experimental evaluation of the energy transferred to EAS components}
The energy fractions of the main EAS components can be estimated using the Yakutsk data. The ionization loss of electrons is measured when detecting the total flux of the Cherenkov light at ground level. The detector disposition of the Yakutsk array is appropriate to measure $Q_{tot}$ in the range above about $Q_{150}=10^7$ m$^{-2}$ as is shown in table~\ref{table:QtotFrac}. In each Q$_{150}$ bin the LDF extrapolation formula~(\ref{Eq.ldf}) is used to calculate the total flux.

\begin{table}
\caption{The fraction of air Cherenkov light, $\Delta_Q$, charged particles, $\Delta_S$, and muons, $\Delta_\mu$, actually measured in EAS events in the Yakutsk array experiment.}
\begin{center}
\begin{tabular}{|c|c|c|c|c|}\hline
 Q$_{150}$, m$^{-2}$ & 10$^6$ & 10$^7$ & 10$^8$ & 10$^9$ \\ \hline
$\Delta_Q, \%$    & 50 & 70 & 90 & 85 \\
$\Delta_S, \%$    &  - &  - & 14 & 13 \\
$\Delta_\mu, \%$  &  - &  - & 65 & 65 \\\hline
\end{tabular}
\end{center}
\label{table:QtotFrac}
\end{table}

Conversion of the measured Q$_{tot}$ to $E_i$ is carried out along equation~(\ref{Eq:Ei2Qtot}) with parameters relevant to the particular observation period, taking into account detector calibration and atmospheric extinction of light.

Other portions of the energy carried out by electromagnetic and muonic components beyond sea level is evaluated via the total number of electrons:
\[
E_g=\epsilon_0 N_e\lambda_e/t_0,
\]
where $\epsilon_0, t_0$ are the critical energy and radiation length of electrons in air; attenuation length $\lambda_e$ is derived from zenith angle dependence of $N_e$~\cite{Prav04}; and muons measured at the ground level:
\[
E_\mu=N_\mu(E>1\:GeV)\overline{E_\mu},
\]
where the average energy of muons $\overline{E_\mu} $ is taken from the MSU array data~\cite{Khrenov}.

Residuary energy fractions transferred to neutrinos $E_\nu$, nucleons $E_h$ etc., unmeasurable with this array, are estimated using computational modeling~\cite{JETP07}. The resulting apportioning of the primary energy $E_0$ is given in table~\ref{table:Balance}.

\begin{table}
\caption{The primary energy fractions of EAS components: ionization loss of electrons in the atmosphere, $E_i$, and in the ground, $E_g$; energy of muons at the ground level, $E_\mu$; energy of EAS components unobservable at the Yakutsk array, $E_{unobs}$. $E_0=E_i+E_g+E_\mu+E_{unobs}$. $\theta=0^0$.}
\begin{center}
\begin{tabular}{|l|c|c|c|}
\hline
Energy deposit & \multicolumn{2}{c|}{$E_0$ fraction, \%}   & Experimental\\\cline{2-3}
channel        & $E_0=10^{18}$ eV &  $E_0=10^{19}$ eV & uncertainty, \%\\
\hline
         $E_i$ &               80 &    77 &                         30 \\
         $E_g$ &                9 &    15 &                         60 \\
       $E_\mu$ &                6 &     4 &                         10 \\
  $E_{unobs}$  &                5 &     4 &                         20 \\
\hline
\end{tabular}
\end{center}
\label{table:Balance}
\end{table}

\begin{figure}
\center{\includegraphics[width=0.6\columnwidth]{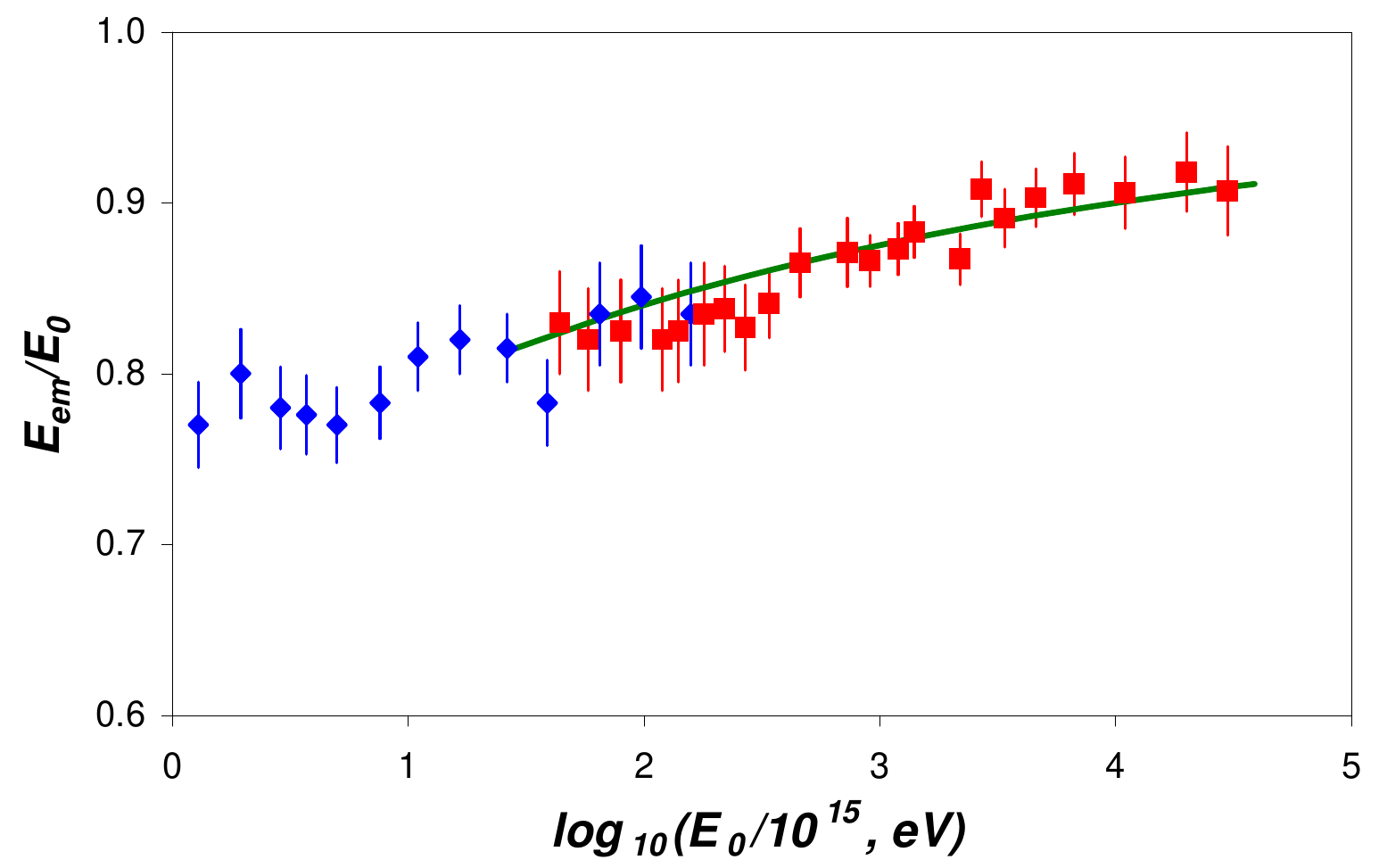}}
\caption{The electromagnetic component energy estimation from the Yakutsk
array data: $C_1$ subset (squares), $C_2$ subset (rhombuses); solid curve is the result of CORSIKA/QGSJET based estimation.}
\label{Fig:Eem}
\end{figure}

The energy fraction carried by electromagnetic component ($E_{em}=E_i+E_g$) appears to be the basic contribution to the total energy of the shower, and its energy dependence (measured with Cherenkov light detectors + scintillators of the Yakutsk array~\cite{Eem}) is illustrated in figure~\ref{Fig:Eem} in comparison with CORSIKA/QGSJET estimation~\cite{Song}.

Due to the air Cherenkov total light flux and the electron and muon number which are experimental values measured at ground level, only about 5\% of the primary energy in the interval $E_0\in(10^{18},10^{19})$ eV is calculated by using model assumptions. So we consider the energy estimation used to be model-independent within these bounds.

\begin{figure}
\center{\includegraphics[width=0.6\columnwidth]{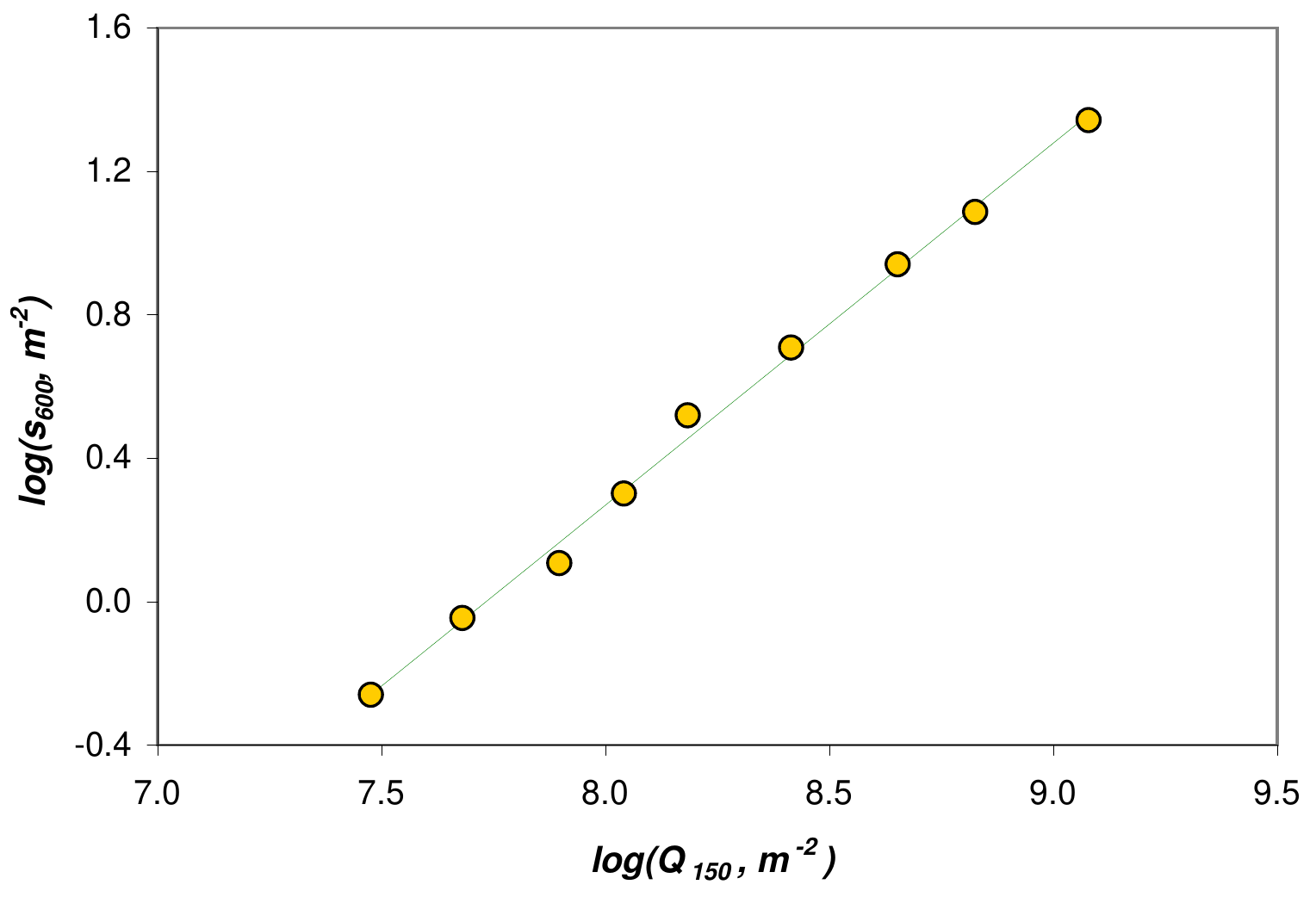}}
\caption{The correlation between charged particle density at 600 m from the core, $S_{600}$, and air Cherenkov light intensity at 150 m, $Q_{150}$, measured in the same showers with $\theta<15^0$.}
\label{Fig:Correl}
\end{figure}

Moonless nights, when air Cherenkov light measurements are possible, constitute $\sim10\%$ of the observation period. In order to evaluate the primary energy of the bulk of showers, the correlation
$$
S_{600}=1.56\times 10^{-8}Q_{150}^{1.01}
$$
is used between the charged particle density at 600 m from the shower core, $S_{600}$, and the light intensity at 150 m from the core (figure~\ref{Fig:Correl}) which, in turn, is related to the total flux of the Cherenkov light in the atmosphere $Q_{tot}=a_1 Q_{150}^{b_1}$, parameters are given in first columns of the table~\ref{T:E2Q150}. Using derived relation between the primary energy and $Q_{tot}$ (equation \ref{Eq:Ei2Qtot}) the final formula was found to link $Q_{150}$ with $E_0$ in the interval $\theta<15^0$~\cite{JETP07}:
\begin{equation}
E_0=a_2 Q_{150}^{b_2},
\label{Eq:Q150toE0}
\end{equation}
where the numerical values of $a_2,b_2$ are summarized in the last two columns of table~\ref{T:E2Q150} for different $Q_{150}$ intervals.

\begin{table}
\caption{Parameters of the relation between $Q_{150}$ and the total flux of air Cherenkov light and the energy of the primary particle initiating EAS.}
\begin{center}
\begin{tabular}{|c|c|c|c|c|}
\hline
$Q_{150}$, m$^{-2}$ & $a_1$ & $b_1$ & $a_2$ & $b_2$\\
\hline
 $<10^6$ & $2.39\times10^6$ & 0.90 & $(6.87\pm1.44)\times10^{10}$ & $0.87\pm0.02$\\
$10^6-10^8$ & $1.33\times10^6$ & 0.94 & $(3.78\pm0.72)\times10^{10}$ & $0.91\pm0.02$\\
$10^8-10^9$ & $7.91\times10^5$ & 0.97 & $(2.90\pm0.29)\times10^{10}$ & $0.93\pm0.02$\\
$>10^9$ & $5.75\times10^5$ & 0.98 & $(5.59\pm0.56)\times10^9$ & $1.01\pm0.02$\\
\hline\end{tabular}
\end{center}
\label{T:E2Q150}\end{table}

\begin{figure}
\center{\includegraphics[width=0.6\columnwidth]{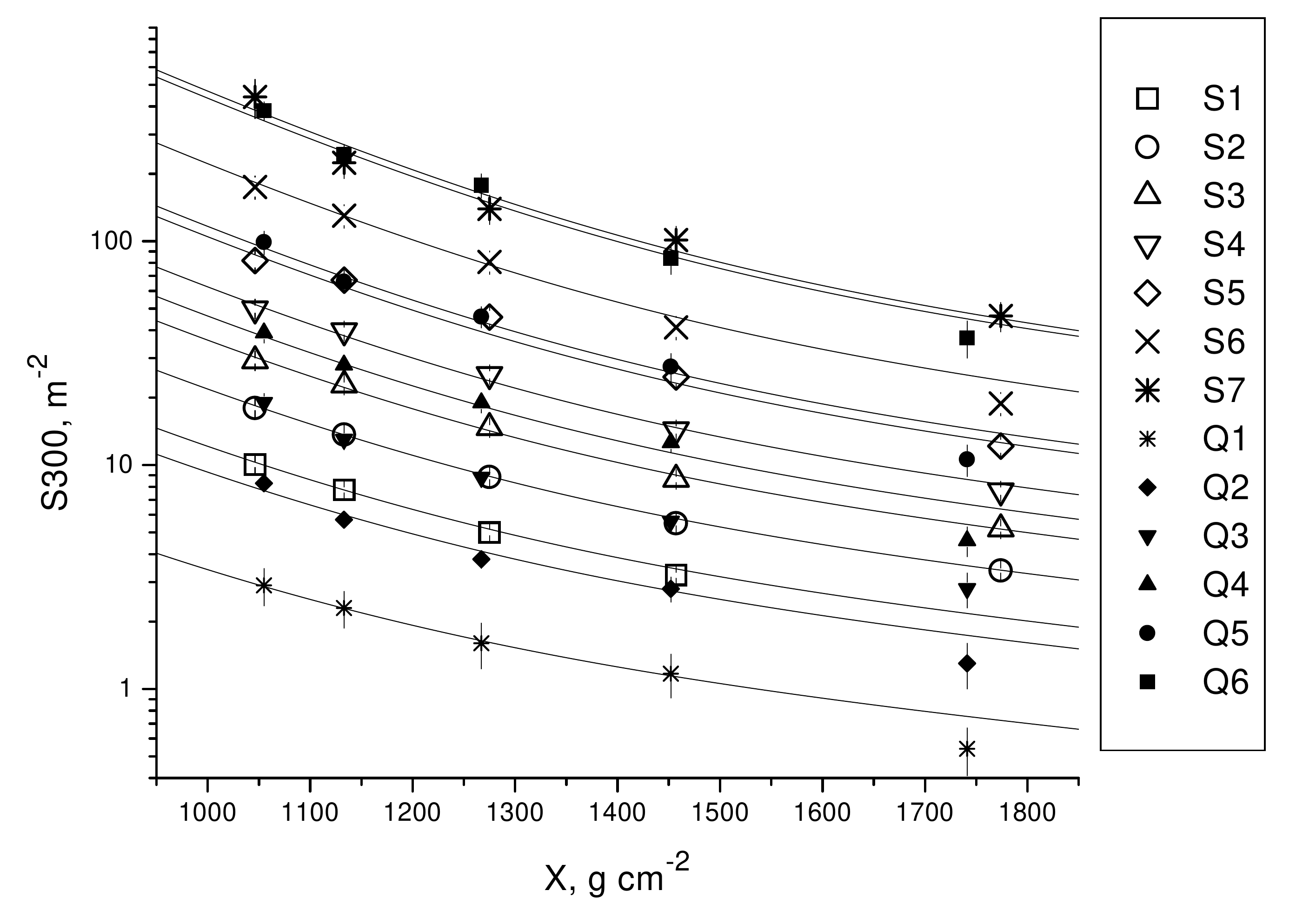}}
\caption{$S_{300}$ as a function of $x=1020/\cos\theta$ for different CR intensities. Open symbols ($S_i$) are equi-intensity method results, filled ones ($Q_i$) are derived fixing $Q_{400}$.}
\label{Fig:EquiCuts}
\end{figure}

The observed densities $S_{300}/S_{600}$ at various zenith angles are connected to the 'vertical' one ($\theta=0^0$) along attenuation curve \cite{Iv01}. In order to measure the attenuation length of these densities for fixed energy, we have used two different methods - well-known equi-intensity cut method, and fixing the Cherenkov light intensity at 400 m from the core as the equivalent of the primary energy, taking into account the light absorption in the atmosphere. In figure~\ref{Fig:EquiCuts} the results are given. Experimental points are consistent with each other for the two methods used and can be described by the sum of two components - a soft component (electrons, attenuation length $\lambda_e=200\: g/cm^2$) and a hard component (muons, $\lambda_\mu=1000\: g/cm^2$)~\cite{Prav04}:
\begin{eqnarray}
S_{300}(\theta)=S_{300}(0^0)((1-\beta_{300})exp(\frac{x_0-x}{\lambda_e})+
\beta_{300}exp(\frac{x_0-x}{\lambda_\mu})),
\end{eqnarray}
where $\beta_{300}$ is the hard-component fraction. Attenuation curve for $S_{600}$ is the same but $\beta$ is different:
\begin{eqnarray}
\beta_{300}=(0.563\pm 0.032)S_{300}(0^0)^{-0.185\pm 0.02},\nonumber\\
\beta_{600}=(0.62\pm 0.006)S_{600}(0^0)^{-0.076\pm 0.03}.
\end{eqnarray}

Experimental uncertainties when estimating the EAS component energies are summarized in the last column of the table~\ref{table:Balance}. The main contribution arise from $\delta E_i$ which is governed by uncertainties in the atmospheric transparency (15\%), detector calibration (21\%) and the total light flux measurement
(15\%). Errors in estimation of $N_e, \lambda_e, N_\mu$ determine the next two items (for ionization loss in the ground and $\delta E_\mu$). Resultant uncertainty in energy estimation is the sum of all errors weighed with the second/third columns of the table~\ref{table:Balance}: $\delta E_0 \sim 32\%$~\cite{JETP07}. Extra 20\% are added due to a $Q_{150}$-$S_{600}$ conversion uncertainty.

To illustrate the energy estimation method used, four showers detected at the Yakutsk array in the range $E_0\geq10^{20}$ eV, $\theta<60^0$, and axes within the array area, are given in table~\ref{Table:UHECRs} (one event is added slightly below the threshold because the energy estimation error is larger than the tiny difference).

\begin{table}
\caption{The highest energy EAS events detected with the Yakutsk array}
\begin{center}
\begin{tabular}{@{}|c|c|c|c|c|c|}
\hline
  Date & $\theta^0$&$\log E_0$&$\delta E_0$,\%& $b^0$& $l^0$ \\
\hline
18.02.04 &   47.7   &  20.16  &      42      & 16.3 & 140.2 \\
07.05.89 &   58.7   &  20.14  &      46      &  2.7 & 161.6 \\
21.12.77 &   46.0   &  20.01  &      40      & 50.0 & 220.6 \\
15.02.78 &    9.6   &  19.99  &      32      & 15.5 & 102.0 \\
\hline
\end{tabular}\\[2pt]
\end{center}
\label{Table:UHECRs}
\end{table}

%%%%%%%%%%%%%%%%%%%%%%%%%%%%%%%%%%%%%%%%%%%%%%%%%%%%%%%%%%%%%%%%%%%%%%%%%%%%%%%%%%%%%%%%%%%%%%%%%%%%%%%%%%

\section{The energy spectrum of cosmic rays derived from air Cherenkov light measurements}\label{Sctn:8}
The Cherenkov light detector subsets of the Yakutsk array give us the opportunity to reconstruct the energy spectrum of cosmic rays in the energy range from $E_0\sim 10^{15}$ to $6\times 10^{19}$ eV. The total number of electrons and muons measured at the ground level are used to estimate the additional energy fractions carried by EAS components but the final relation~(\ref{Eq:Q150toE0}) comprises the light intensity alone.

The intensity of CRs has been evaluated using the number of EAS events derived from the showers detected with PMTs, and aperture $S_{eff}T\Omega$ of the sub-arrays in a particular energy interval, where the array area bounded by the perimeter, $S_{eff}$, depends on the primary energy and zenith angle; $\Omega$ is the acceptance solid angle; and $T$ is the sum of observation periods. Under the condition of the array configuration changing due to rearrangements and non-active detectors, $S_{eff}$ has to be calculated for a given period using the Monte Carlo technique.

\begin{figure}
\center{\includegraphics[width=0.6\columnwidth]{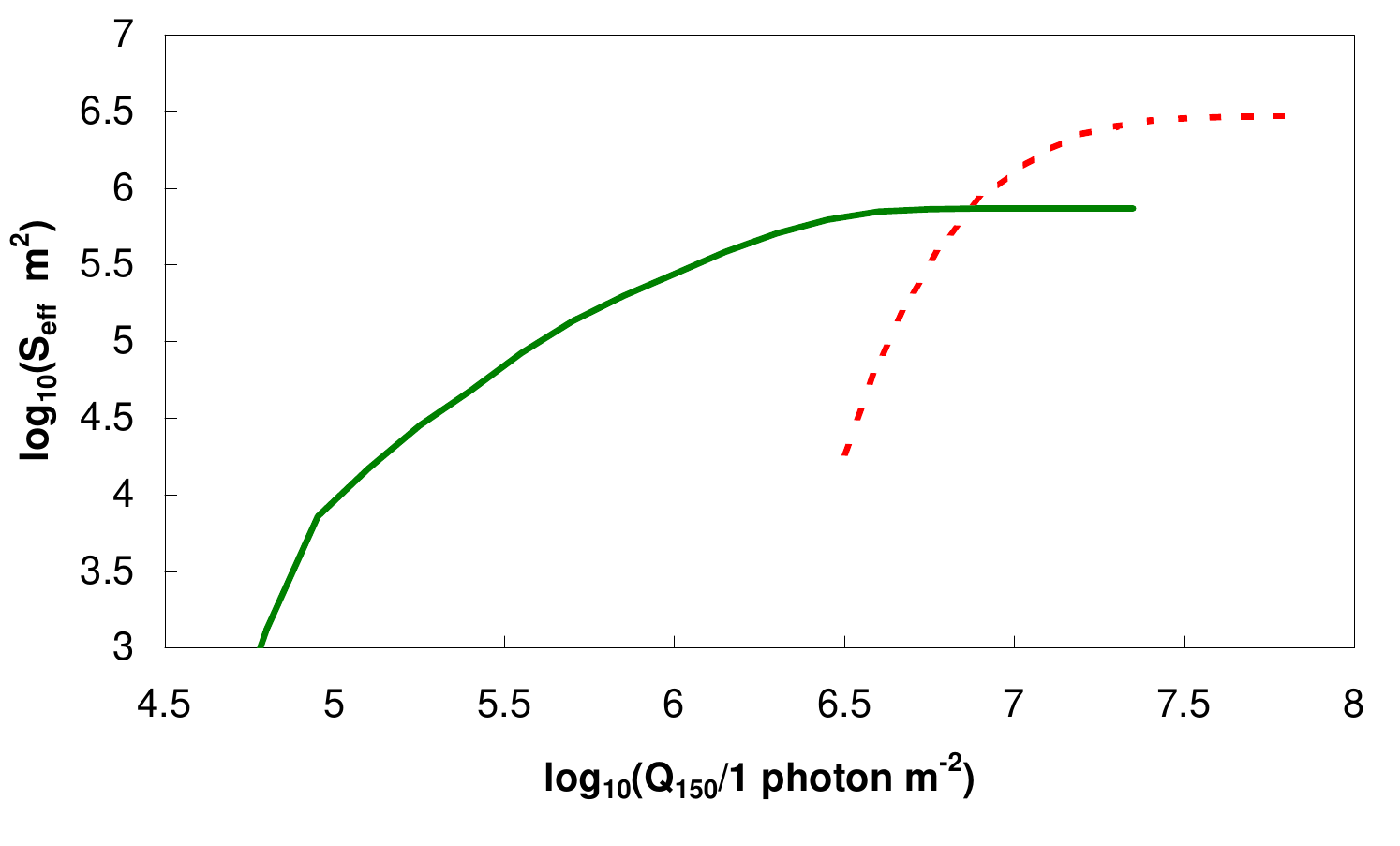}}
\caption{Acceptance area of air Cherenkov light detector subsets of the Yakutsk array. The solid curve is for autonomous, $C_2$, and the dotted one for the medium, $C_1$, sub-array area.}
\label{Fig:Seff}
\end{figure}

The acceptance area has been simulated as a function of $Q_{150}$ averaged in the zenith angle interval ($0^0,30^0$) for the two Cherenkov light detector subsets shown in figure~\ref{Fig:Map}. A lateral distribution fit accord to Eq.~(\ref{Eq.ldf}) is used together with instrumental and statistical errors (Gaussian with 25\% relative deviation) to model the trigger of each detector subset with 100000 fake showers. In the case of the autonomous sub-array, the Cherenkov trigger is simulated, while in the $C_1$ case the double trigger for scintillator and PMT signals has been modelled. The average ($S_{eff}=S_0 n_{triggered}/10^4$, where $S_0$ is the array area inside the perimeter; $n_{triggered}$ is triggered number of events) is shown in Figure~\ref{Fig:Seff} versus the parameter $Q_{150}$. Corrections for inoperative detectors are not shown here.

The shower data gathered after the latest array re-configuration were used to work out the spectrum. Namely, 1993-2007 for the medium $C_1$ subset, and 1995-2007 for autonomous subset. In order to evaluate the intensity of the primary flux we have collected data during observation periods with a light extinction better than 0.65 and shower axes within area of the corresponding sub-array.

\begin{figure}
\center{\includegraphics[width=0.6\columnwidth]{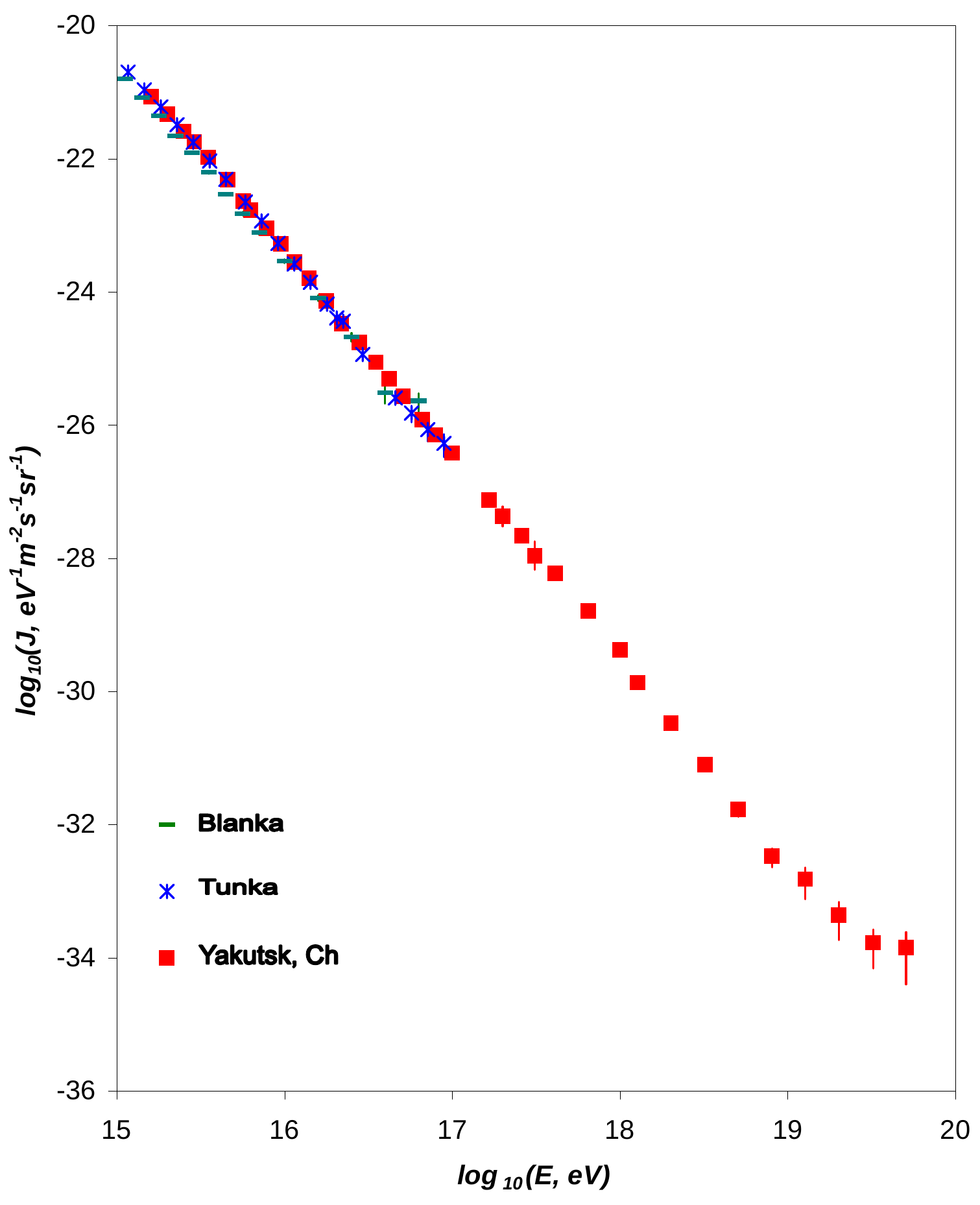}}
\caption{Cosmic ray flux measured by arrays equipped with air Cherenkov light detectors. Vertical bars indicate statistical errors.}
\label{Fig:Spectra08}
\end{figure}

The resulting differential all-particle spectrum of cosmic rays is shown in figure~\ref{Fig:Spectra08} in comparison with the data from other Cherenkov detector arrays, namely, BLANKA~\cite{Blanca} and Tunka~\cite{Tunka}. The present data (given in tabular form in~\ref{app-c}: tables \ref{Table:AutoData}, \ref{Table:MainData}) exhibit the spectrum irregularity near $E\sim 10^{19}$ eV, the 'ankle', seen by all arrays in the area~\cite{Nagano}; at lower energies the results of all three arrays are compatible with a 'knee' at $E\sim3\times10^{15}$ eV revealed in the pioneering works of the MSU array~\cite{Khristi}. Below $E=10^{18}$ eV there is a transition region between the two subsets, $C_1,C_2$, so the 'second knee' visible here may be due to the data sewed together.

Hereinafter, two energy regions below and above $E=10^{18}$ eV are analyzed separately in order to consider the two irregularities in the spectrum.

\begin{figure}
\includegraphics[width=0.47\textwidth]{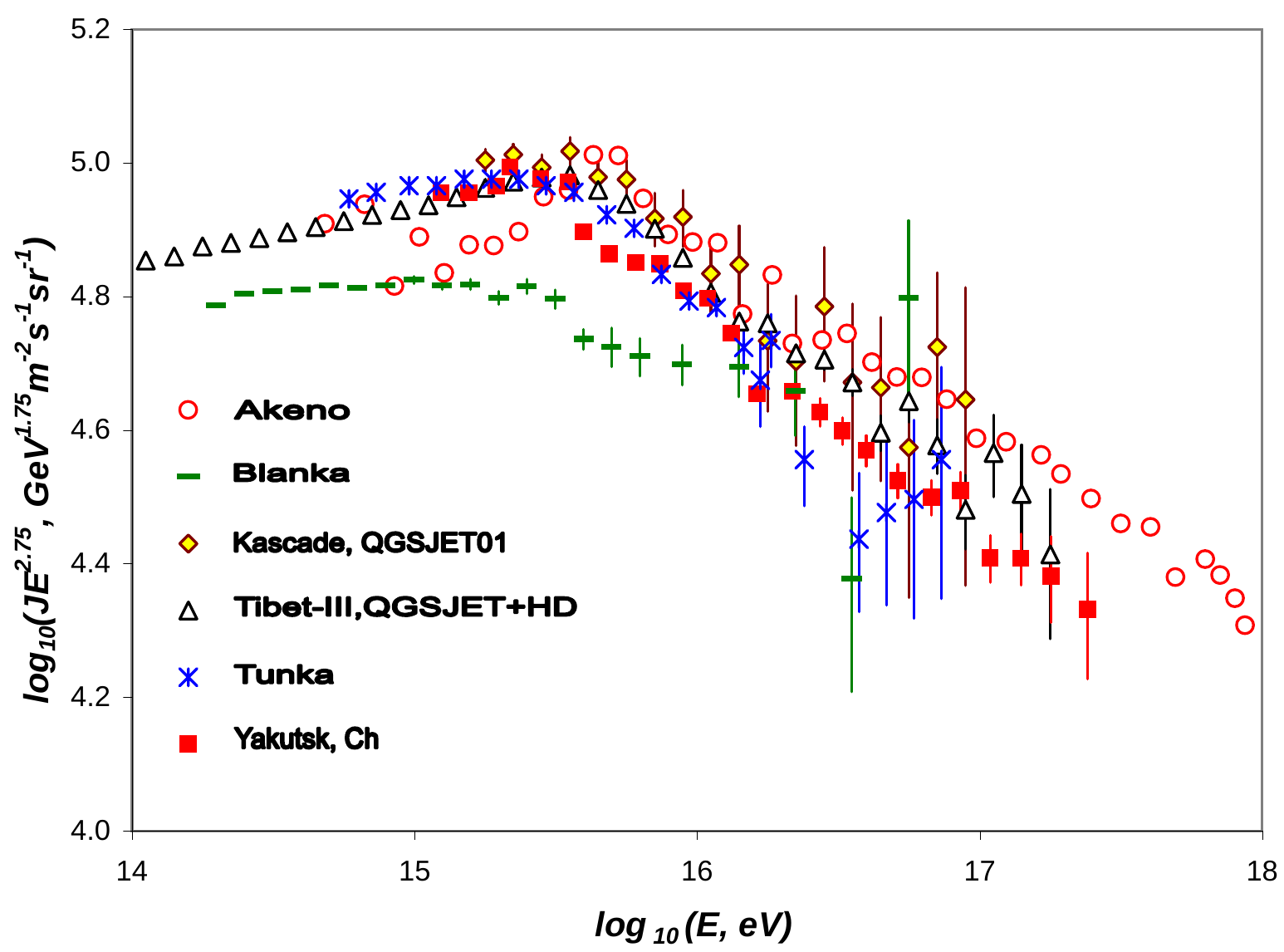}
\hskip0.05\textwidth
\includegraphics[width=0.48\textwidth]{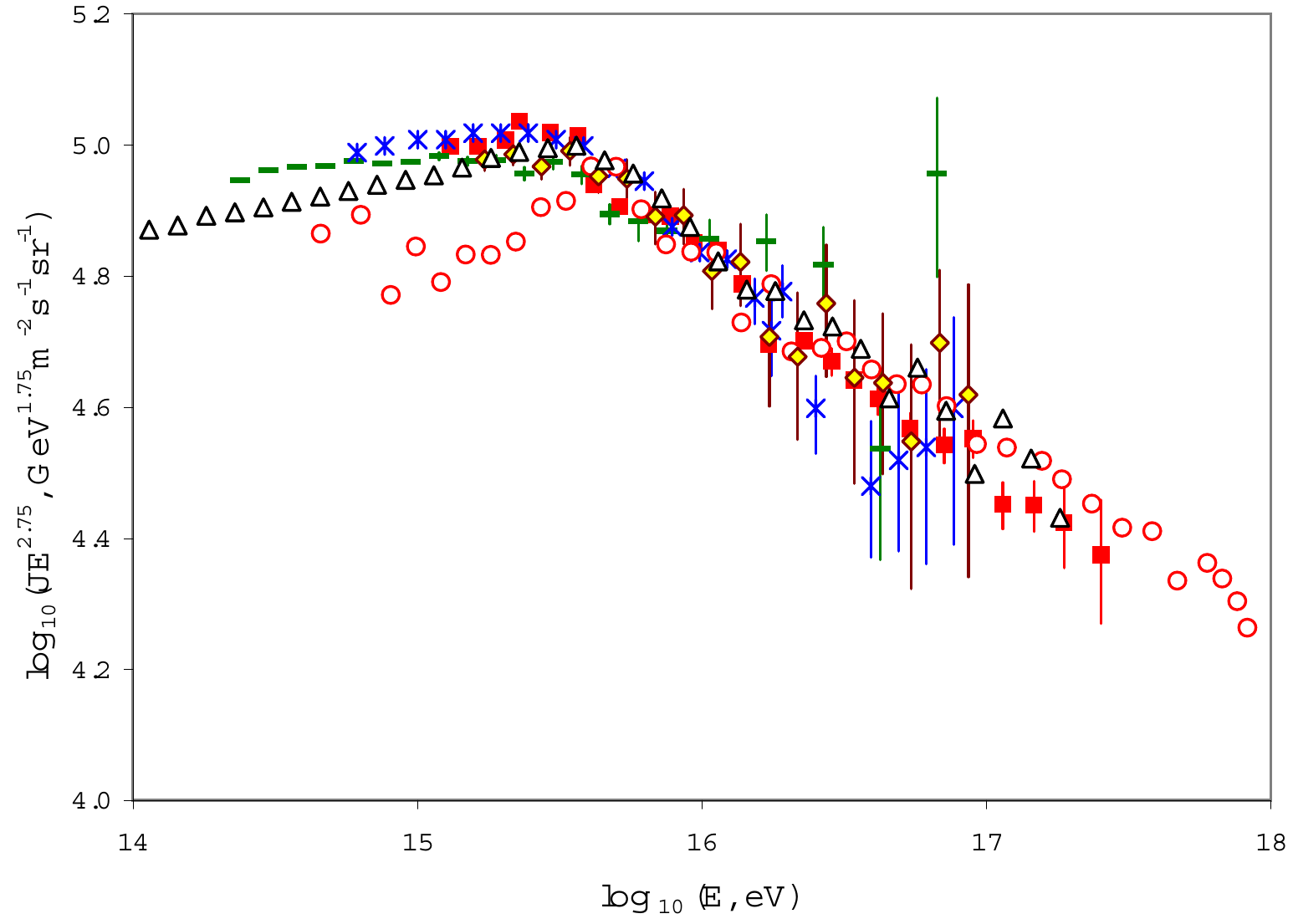}
\caption{The energy spectrum of cosmic rays around the knee region. Observational data are from arrays equipped with Cherenkov light and scintillation detectors. Two panels show the spectra before (left) and after (right) energy corrections applied to the data, as described in the text.}
\label{Fig:KneeR2}
\end{figure}

The region below $10^{18}$ eV is shown in detail in figure~\ref{Fig:KneeR2} where CR intensity is multiplied by $E^{2.75}$ in order to emphasize the spectrum irregularity. In addition to spectra around the knee measured by Cherenkov detector arrays (BLANKA, Tunka and Yakutsk taken from the previous Figure), the data of  scintillation detector arrays (Akeno~\cite{Akeno}, KASCADE~\cite{Kascade} and Tibet~\cite{Tibet}) are given for comparison. Due to air Cherenkov light signal proportional to the number of electrons in the shower, no difference is expected in the spectrum shape measured with scintillator or PMT arrays.
Measured CR intensities are corrected in the case of Cherenkov detector arrays along the algorithm in \ref{app-b}. Integral energy spectrum indices below and above the knee are assumed to be 1.67 and 2.1, respectively, as measured by the Tibet-III array, while energy evaluation errors are estimated as 0.12 (Blanka), 0.2 (Tunka) and 0.25 (Yakutsk). Original intensities from Akeno, KASCADE and Tibet arrays are not changed because the energy/intensity reconstruction procedures from $N_e,N_\mu$ measurements include the conversion factor (\ref{Eq:NNK}) needed.

For the reconstructed energy and intensity there is an interaction model/primary composition dependence in all the data from arrays. For instance, the Tibet-III results indicate a $\sim20\%$ systematic error due to chemical composition and $\sim10\%$ discrepancy between QGSJET01c and SIBYLL2.1 interaction models below $E=10^{16}$ eV. The uncertainty may by much worse if to use interaction models not carefully tuned to the measured EAS observables.

We have used energy correction factors to compare the measured differential spectra around the knee. The intensities ($J\times E^{2.75}$) shown in the left panel of the figure~\ref{Fig:KneeR2} were shifted by factors 1.2, 1.05 and 1.05 in Blanka, Tunka and Yakutsk cases, respectively; other three results were shifted along energy correction factors 0.95 (Akeno), 0.97 (KASCADE) and 1.02 (Tibet). The resultant spectra are plotted in the right panel of the figure~\ref{Fig:KneeR2}.
All data exhibit the knee approximately in the same energy interval while the intensity/energy estimates are different especially in the Akeno and BLANCA flux data.

\begin{figure}
\center{\includegraphics[width=0.6\columnwidth]{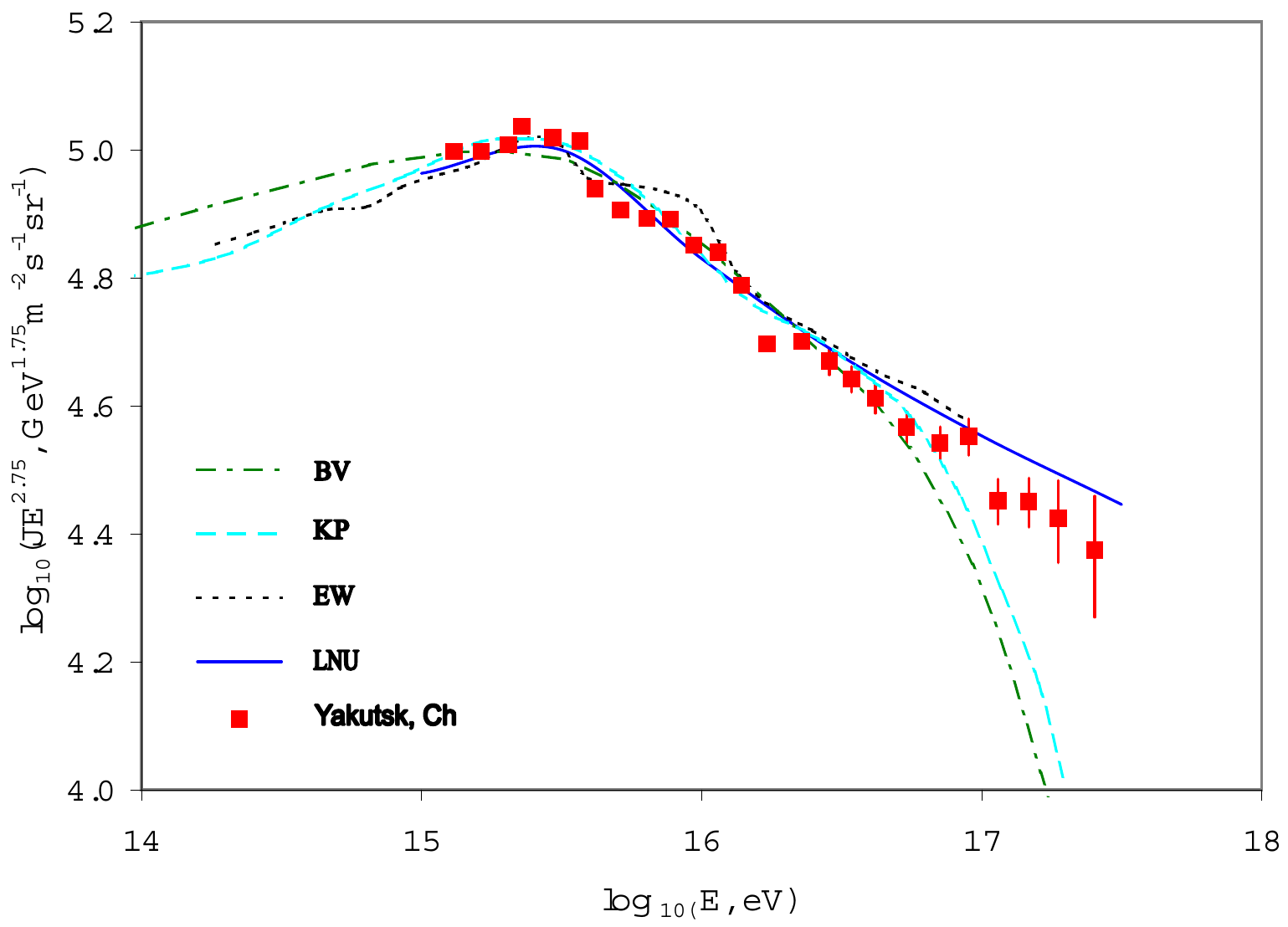}}
\caption{Our energy spectrum and the results of galactic cosmic ray simulations. Curves: SNR acceleration model~\cite{Berezhko} (BV); diffusion model~\cite{Kalmykov} (KP); single source model~\cite{SSM1} (EW) and anomalous diffusion model~\cite{Lagutin} (LNU).}
\label{Fig:ModelSpectra}
\end{figure}

As examples of model calculations, four predicted spectra are shown together with the Cherenkov light data of the Yakutsk array in figure~\ref{Fig:ModelSpectra}. Only the shape of the all-particle spectra can be compared with the data because of the free parameters in models - the intensity of CRs and, to a lesser degree, a knee position due to the magnetic field uncertainty in the sources and in the interstellar medium.

All the model spectra are compatible with experiment, especially, in view of the dispersion in the results of arrays (e.g. figure~\ref{Fig:KneeR2}) which is greater then the shape variation due to models. At energy $E>0.1$ EeV the anomalous diffusion model spectrum is harder then that observed in Yakutsk, but the points are within experimental errors. The lack of CRs in SNR acceleration and diffusion models above energy $0.1$ EeV can be filled up by the extragalactic component.

There is seen to be a hint of the fine structure in the energy spectrum measured with the Yakutsk array. Although it is possible that the undulations around the knee are caused by the instrumental errors\footnote{vertical bars show statistical errors only}, but on the other hand, there are 40 size spectra and 5 Cherenkov light spectra measured before 2005, which demonstrate the second excess ('peak') at $\lg E=\lg E_{knee}+0.6$ besides the knee itself~\cite{SSM2}. The second peak in our data is approximately at this energy. However, more Cherenkov light data are needed to scrutinize the subject. Our autonomous sub-array is able to supply with a sufficient sample of EAS events within the next few years.

Recently, Lagutin et al.~\cite{LagutinSSM} examined the contribution of a nearby SNR-type source to the energy spectrum of CRs produced in anomalous diffusion model. They found several sequential peaks caused by H, He and CNO nuclei around the knee in all-particle spectrum, confirming results of the single source model~\cite{SSM1} in the case of the 'background' anomalous diffusion model.

Our conclusion concerning the part of the energy spectrum below $10^{18}$ eV is that while all four models considered are compatible with our measurements in the knee region, e.g. the intensities below and above the knee, the only model able to describe the fine structure in the spectrum is the single source model. So this model combining a recent nearby SNR with the background (anomalous) diffusion of CRs in Galaxy is the best fit for the Yakutsk array data.

When comparing the upper half of the spectrum measured for $E>10^{18}$ eV, with the energy spectra observed by giant EAS arrays, including that of the Yakutsk array, the somewhat different energy estimation has to be taken into account~\cite{Mono,Prav04,CRIS}.
Due to the comparatively small acceptance area of the Cherenkov detector subsets (figure~\ref{Fig:Seff}), our data are reliable up to $\sim10^{19}$ eV; this includes the ankle region. Another purpose is to compare UHECR intensities measured with different techniques.

In the table~\ref{Table:RJ} are given the energy estimation errors of the AGASA~\cite{AGASASpectrum}, HiRes~\cite{HiRes08}, PAO~\cite{PAO08} and Yakutsk~\cite{JETP07} experiments and the intensity conversion factors calculated as described in \ref{app-b} for energy bins where the index is constant and the variation of instrumental errors is negligible. The integral energy spectrum index of CRs is assumed to be 2.3 and 1.9 below and above the ankle in the spectrum, and $\kappa=4$ at $\lg E>19.8$, along the results of HiRes fitted by the broken power law~\cite{HiRes08}.

\begin{table}
\caption{Energy estimation errors, $\sigma$, and intensity conversion factors, $R_J$, for EAS arrays.}
\begin{center}
\begin{tabular}{lcccc}&&&&\\
\hline
               Array & AGASA & HiRes &  PAO & Yakutsk \\
\hline
        $\sigma$, \% &    25 &    17 &   22 &    32   \\
     $R_J(lgE<18.5)$ &  1.18 &  1.08 & 1.14 &  1.31   \\
$R_J(18.5<lgE<19.8)$ &  1.12 &  1.05 & 1.09 &  1.20   \\
     $R_J(lgE>19.8)$ &  1.65 &  1.26 & 1.47 &  2.27   \\
\hline
\end{tabular}
\end{center}
\label{Table:RJ}
\end{table}

\begin{figure}
\includegraphics[width=0.48\textwidth]{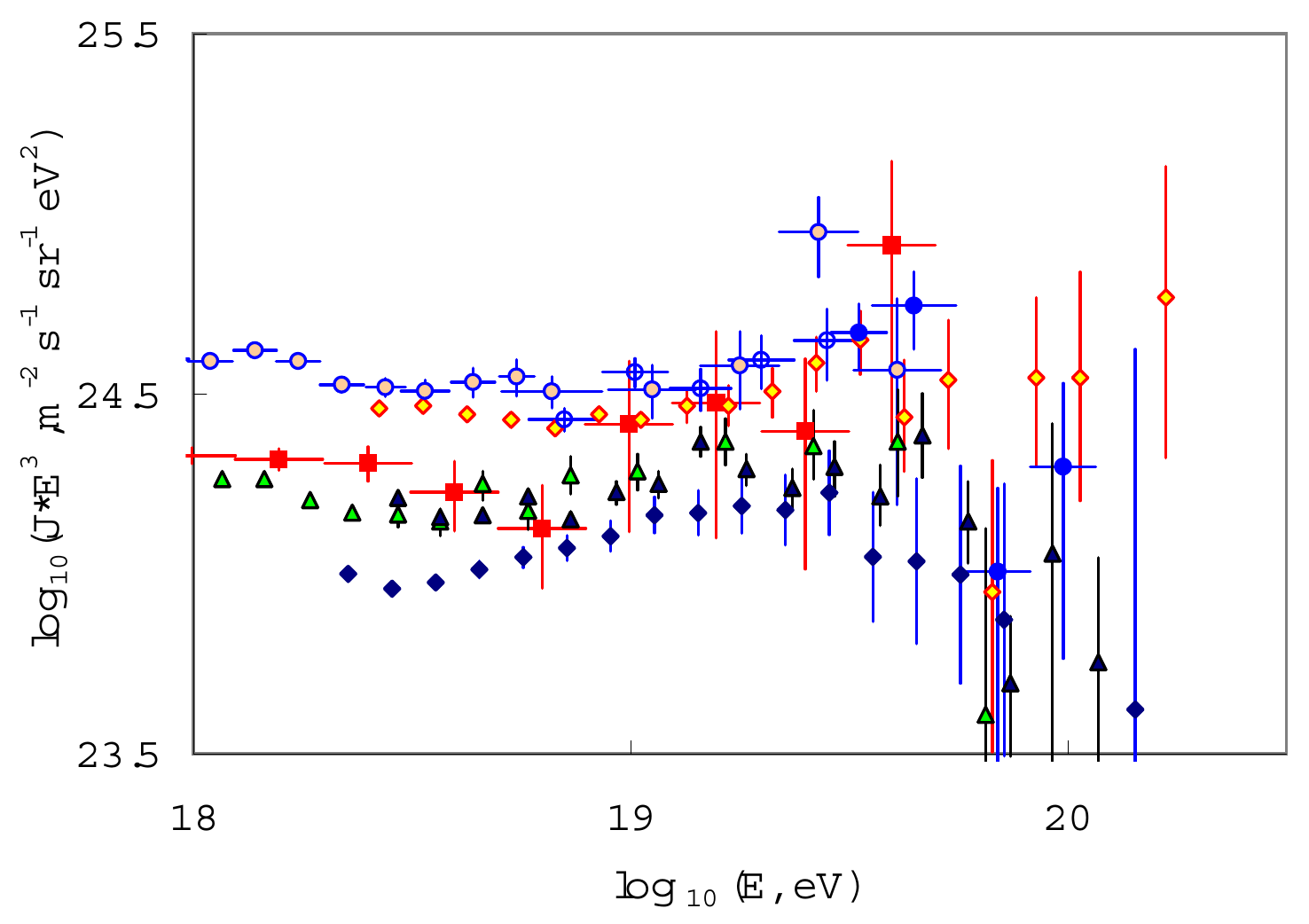}
\hskip0.04\textwidth
\includegraphics[width=0.48\textwidth]{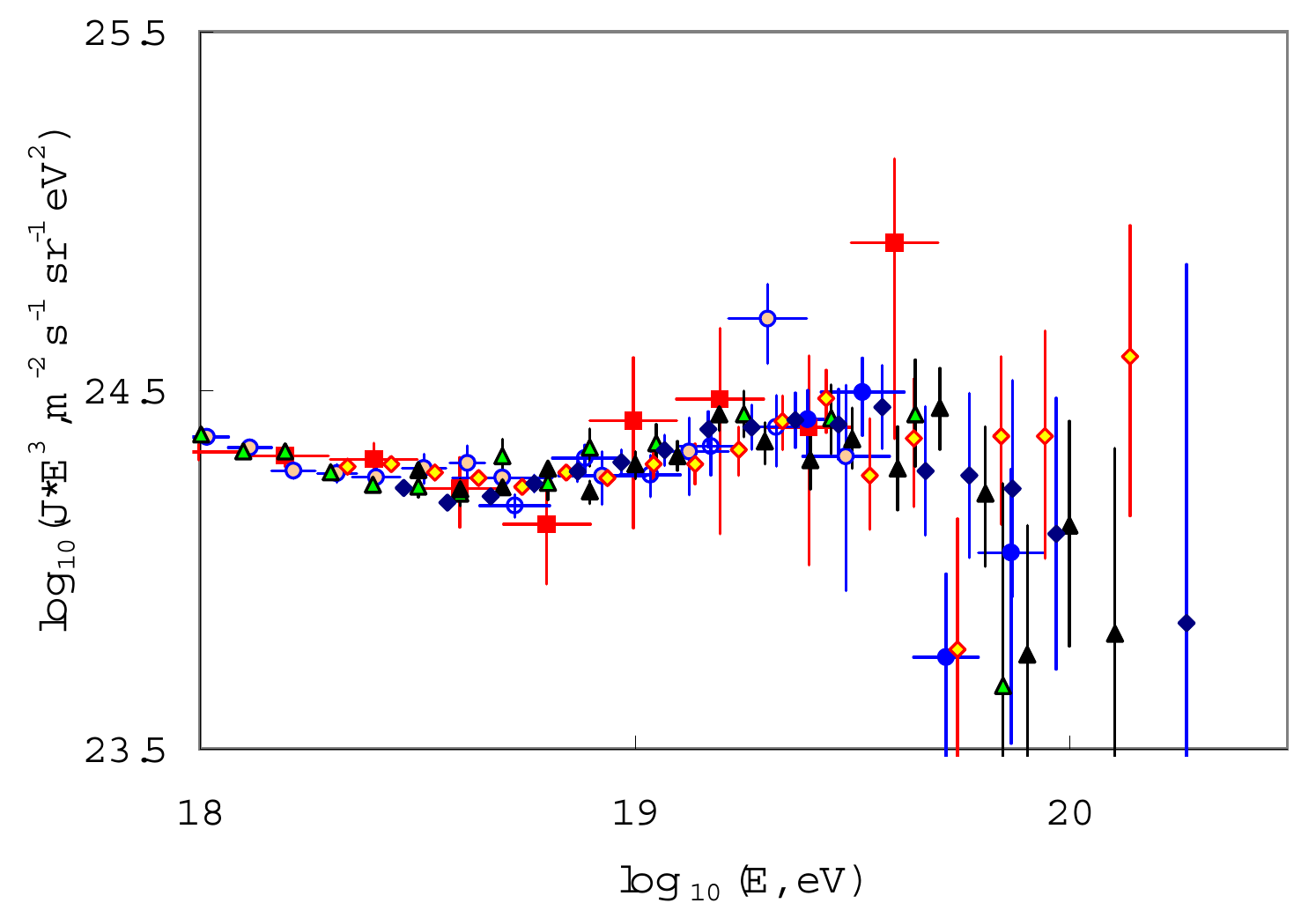}
\caption{UHECR energy spectra measured by AGASA (open rhombuses)~\cite{AGASASpectrum}, HiRes I, II (open and filled triangles)~\cite{HiRes08}, PAO (filled rhombuses)~\cite{PAO08}, the Yakutsk array scintillation detectors (open, crossed and filled circles)~\cite{Prav04}, and air Cherenkov light detectors (squares)~\cite{Merida}. Statistical errors are shown by the vertical bars, while horizontal bars indicate energy bins. The spectra are given before (left panel) and after (right panel) energy corrections applied to the data.}
\label{Fig:SpectraCorrJ}
\end{figure}

Observed UHECR spectra are given in the left panel of the figure~\ref{Fig:SpectraCorrJ} with the intensity correction factors applied: observed intensities are decreased by $R_J$ and spectra are displaced over $0.434\sigma$ along $\lg E$.\footnote{This may be unwarranted in the PAO case due to~\cite{Carvalho}.}
This is preliminary, crude procedure to reconstruct the spectrum breaks, but for the purpose of the intensity comparison it may be sufficient.

Two spectra of the Yakutsk array are compatible within errors above $10^{19}$ eV but diverge at lower energies. The possible reason of a discrepancy can be systematic errors in primary energy and CR intensity estimation near the scintillator threshold of the trigger-500. The work is in progress to surmount the divergence.

Comparison of UHECR spectra measured using different detectors and energy reconstruction methods infer the existence of systematic differences between resultant energies obtained at the arrays. The average values, $\overline{\hat{E}}_i$, should be corrected, too.\footnote{for the given primary energy E} While the true primary energy is unknown, cross calibration of energy estimation methods can be carried out adjusting correction factors, $R_E$, for the pairs of observed spectra, converging it together. Then the resulting spread of factors elucidates the confidence interval for the CR energy estimated.

Table~\ref{Table:iCalibr} demonstrates the variety of correction factors for the energy estimation methods of the different groups.

\begin{table}
\caption{Correction factors to energy scales for the pairs of EAS arrays/detectors, $R_E$, averaged in the region $E>10^{18}$ eV.}
\begin{center}
\begin{tabular}{llllll}&&&&&\\
\hline
        & AGASA & HiRes & PAO & $Y_{Sc}$ & $Y_{Ch}$ \\
\hline
AGASA    &    1 & 0.75 & 0.63 & 1.05 & 0.82 \\
HiRes    & 1.33 &    1 & 0.85 & 1.40 & 1.08 \\
PAO      & 1.6  & 1.2  &    1 & 1.70 & 1.30 \\
$Y_{Sc}$ & 0.91 & 0.71 & 0.6  &    1 & 0.75 \\
$Y_{Ch}$ & 1.22 & 0.93 & 0.80 & 1.33 &   1  \\
\hline
\end{tabular}
\end{center}
\label{Table:iCalibr}
\end{table}

To illustrate the result of corrections $R_j\times R_E$ applied to estimated energies and intensities, the measured spectra are re-plotted in the right panel of figure~\ref{Fig:SpectraCorrJ}. Energy scale factors are used here (arbitrarily, but preferably close to the median) from the last column of table~\ref{Table:iCalibr}, although any other column may be used as well.

\begin{figure}
\center{\includegraphics[width=0.6\columnwidth]{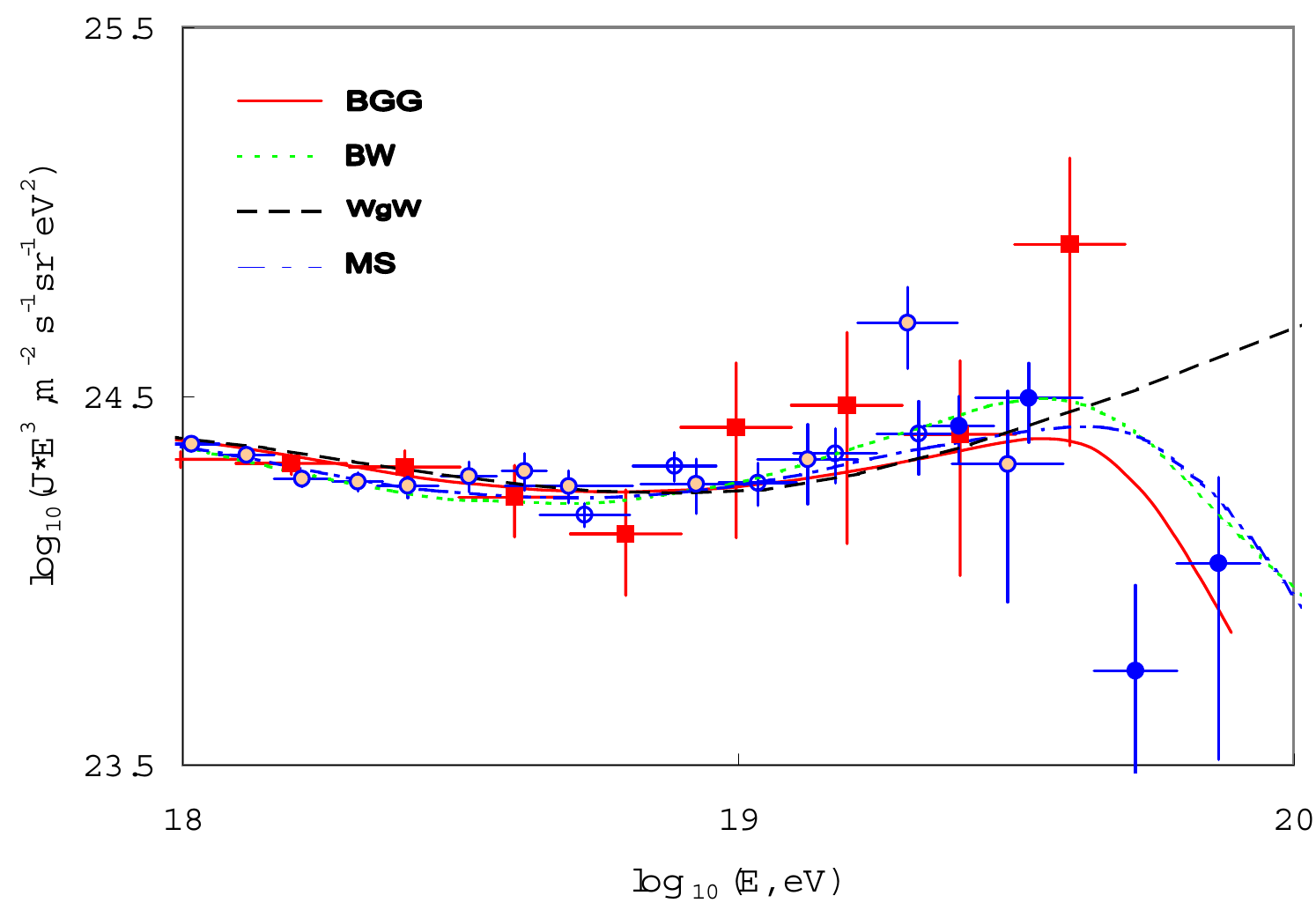}}
\caption{The Yakutsk array data from the previous figure (with energy correction) in comparison with the results of UHECR propagation models given by curves: Berezinsky, Gazizov \& Grigorieva (BGG); Bahcall \& Waxman (BW); Wibig \& Wolfendale (WgW); De Marco \& Stanev (MS).}
\label{Fig:SpectraFit}
\end{figure}

The shape of the UHECR spectrum measured by all the arrays is compatible within errors, if the energy estimations are calibrated. Namely, the observed position of the ankle and energy threshold of GZK suppression\footnote{except AGASA data} are in satisfactory agreement.

A difference in the energy of EAS primary particle estimated basing on the data of the two subsets of the Yakutsk array detectors - scintillators, $Y_{Sc}$, and air Cherenkov light detectors, $Y_{Ch}$, originates, presumably, in the systematic error of the relation used between the charged particle density at 600 m from the shower core and the light intensity at 150 m from the core.

The UHECR propagation modeling results are illustrated by the four examples in figure~\ref{Fig:SpectraFit} in comparison with the Yakutsk array data.

Bahcall and Waxman gave two-component (Galactic+extra-galactic) model with the shape insensitive to the choice of absolute energy scale~\cite{Bahcall}.

Another approach was used by Berezinsky et al. assuming UHECRs as extragalactic protons from uniformly distributed sources~\cite{Dip07}. The electron-positron pair production in collisions of protons with relic photons results in the energy spectrum of extragalactic CRs with the 'dip' feature in this model.

De Marko and Stanev' model fits the UHECR spectra measured by AGASA and HiRes with different injection spectra at CR sources that are uniformly and homogeneously distributed in the Universe~\cite{MS}. The best fit they found assuming that cosmic rays ($E>10^{19}$ eV) are protons, varying the index, emissivity and cosmological evolution parameters of the injection spectrum, is given in the figure.

Wibig and Wolfendale focus on the ankle in the primary energy spectrum attributing it to the rapid transition from Galactic to Extragalactic component of CRs~\cite{WgW}. The sum of a smoothly falling Galactic spectrum and a power-low EG spectrum\footnote{index -2.37} fitted to the Yakutsk array scintillator data is shown.

All the models used demonstrate GZK suppression\footnote{except~\cite{WgW} which is not intended for} and the ankle features in agreement with the data from arrays. It is not surprising in view of the fact that the source emissivity and the ankle position in the energy scale are free parameters, and GZK effect is embedded in models.

The lack of EAS events above $10^{19}$ eV observed with Cherenkov light detectors hinders in choosing a model of the better fit. The index of the energy spectrum observed in the interval $(10^{19},3\times10^{19})$ eV is $d\lg J/d\lg E=-2.1$ for Cherenkov light detectors data and is $-2.7$ for charged particle detectors data. If we use these values as the confidence bounds then model indices are within the interval. The model of Bahcall and Waxman is closest to our Cherenkov light detectors data and can be considered as preferable among equally matched.

%%%%%%%%%%%%%%%%%%%%%%%%%%%%%%%%%%%%%%%%%%%%%%%%%%%%%%%%%%%%%%%%%%%%%

\section{Conclusions}\label{Sctn:9}
The total flux of air Cherenkov light with subsidiary data on the electron and muon sizes at the ground level is used to estimate the energy of the primary CR particle initiating EAS. The relation between the total flux and ionization loss of electrons in the atmosphere is derived which depends on the extinction of light and $X_{max}$ parameters; the latter is the only parameter to accumulate the interaction model dependence of EAS development in this case.

The independent measurement technique based on the Cherenkov light detectors of the Yakutsk array enabled us to observe the cosmic ray energy spectrum in the range from $E\sim 10^{15}$ eV to $6\times 10^{19}$ eV. Two spectra measured with different detectors of the Yakutsk array - scintillators and Cherenkov light detectors, exhibit an ankle feature below $E=10^{19}$ eV. The autonomous sub-array data confirm the previous observations of the knee at $E\sim 3\times10^{15}$ eV. A comparison of our results with the data of other EAS arrays shows the compatibility of spectra if the energy estimations are corrected.

The energy spectra predicted for several models of galactic CRs demonstrate agreement within experimental errors with our data in the knee region. However, only the single source model describes the fine structure of the spectrum observed attributing it to the contribution of nuclei from a recent nearby supernova. We have chosen this model as the best fit for our data below $10^{18}$ eV.

At the highest energies extragalactic CRs forming a dip, or transition between galactic and extragalactic components above $10^{18}$ eV are thought to be responsible for an ankle detected in the energy spectrum. A comparison of our data with models of these types shows a satisfactory agreement with all of them (the best fit with Bahcall \& Waxman's model), so we cannot distinguish between different scenarios of the ankle formation basing on the energy spectrum measurement alone. There is a need for additional data, presumably, concerning the mass composition of UHECRs, in order to elucidate the origin of an ankle in the energy spectrum.

%%%%%%%%%%%%%%%%%%%%%%%%%%%%%%%%%%%%%%%%%%%%%%%%%%%%%%%%%%%%%%%%%%%%%

\ack
We gratefully acknowledge the contribution to data acquisition and analysis from the Yakutsk array collaboration members. We would like to thank the referees for helpful and productive comments.
The Yakutsk array experiment is supported by the Russian Academy of Sciences. This work is supported by RFBR grants \#06-02-16973, \#05-08-50045.

%%%%%%%%%%%%%%%%%%%%%%%%%%%%%%%%%%%%%%%%%%%%%%%%%%%%%%%%%%%%%%%%%%%%%

\appendix
\section{Energy balance of EAS components}
\label{app-a}
The energy fractions of the EAS primary particle transferred to the shower components can be described on the basis of hadron transport equations. If $E_k,\: (k=N, \pi, \mu\nu, e\gamma)$ is the energy transferred to nucleons, charged pions, muons+neutrinos, electrons+photons, a few cascade parameters determine the ratios between $E_k$ - i.e. the energy balance in the shower. For instance, the transport equation for the charged pions density $\pi (x,E)$ at depth $x$ is:
\begin{eqnarray}
\frac{\partial\pi(x,E)}{\partial x}=-(\frac{1}{\lambda_\pi}+\frac{B_\pi}{x E})\pi(x,E)
+\frac{2}{3\lambda_\pi}\int_{E}^{E_0}\pi(x,U)w_{\pi\pi}(E,U)dU \nonumber\\
+\frac{2}{3\lambda_N}\int_{E}^{E_0}N(x,U)w_{\pi N}(E,U)dU,
\end{eqnarray}
where the interaction mean free paths $\lambda_{\pi},\lambda_N$ are
assumed to be constant; $w_{\pi\pi}(E,U),w_{\pi N}(E,U)$ are the spectra of charged pions produced in pion-air and nucleon-air interactions, can be transformed (integrating over $E^2$) to:
\begin{equation}
\frac{dE_\pi}{dx}=-\frac{E_\pi}{\lambda_\pi}-\frac{B_\pi\pi(x,E>0)}{x}+
\frac{2E_\pi}{3\lambda_\pi}+\frac{2K_N E_N}{3\lambda_N},
\label{eq:Ebalance}\end{equation}
where $E_\pi(x)=\int_{0}^{E_0}\pi(x,E)EdE$; $\pi(x,E>0)=\int_{0}^{E_0}\pi(x,E)dE$; $K_N$ is the nucleon inelasticity assumed to be constant; $E_N=E_0 exp(-K_N
x/\lambda_N)$.

\begin{figure}[t]
\center{\includegraphics[width=0.6\columnwidth]{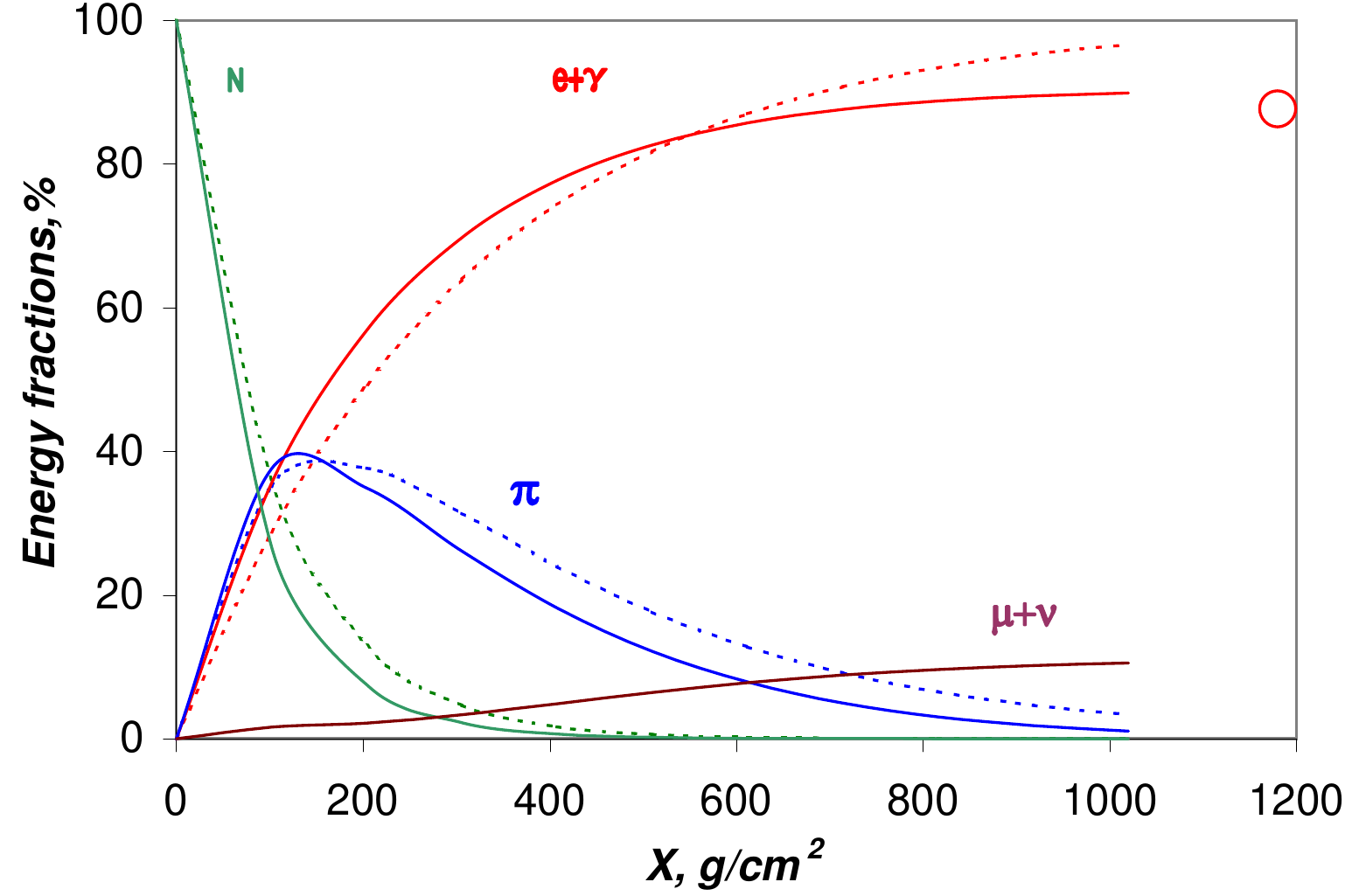}}
\caption{The energy carried in a cascade by nucleons ($N$), charged pions ($\pi$), muons and neutrinos ($\mu+\nu$), electrons and photons ($e+\gamma$). Dashed curves are analytic expressions with $\lambda_i, K_N=const, B_\pi=0$: $E_N=E_0exp(-K_N x/\lambda_N)$; $E_\pi=2/3E_0(1-exp(-K_Nx/\lambda_N))exp(-x/\lambda_\pi/3)$; $E_{e\gamma}=E_0-E_N-E_\pi$. Solid curves are $\delta$-model results: cross sections are supposed rising $\propto 0.08lnE$; $B_\pi=120$ GeV; $K_N=0.5$; $n_s\propto E^{1/4}$; $E_0=10^{18}$ eV. The open circle represents the asymptotic estimation for $E_{e\gamma}$ of the CORSIKA/QGSJET code.}
\label{Fig:Balance}
\end{figure}

In the energy range $E\gg B_\pi$ the only parameters to define the solution are $K_N/\lambda_N$ and $\lambda_\pi$. It means that the energy transferred to charged pions is independent of the spectra of pions produced in nuclear interactions. Hence, in the general case, we can use simple $\delta$-model with the production spectrum $w_{ik}(E,U)=n_s\delta(E-U/n_s)$, where $n_s$ is the multiplicity of secondaries, to balance the components energy in a shower. Because of the net value of $E_{\mu+\nu}/E_0\leq 0.1$, the uncertainty due to the simplified model should be of second order of magnitude.

To summarize, the main model parameters governing the energy balance in the shower are the average inelasticity coefficients, mean free paths, multiplicity of secondaries and the fragmentation rate of the primary nucleus. The influence of other model characteristics such as 'the form of the rapidity distribution of constituent quarks' in collisions is weak.

The exact analytic solution of equations in the case of constant $\lambda_{\pi},\lambda_N, K_N$ and $B_\pi=0$ is shown in figure~\ref{Fig:Balance} together with $\delta$-model results. The letter is demonstrating the influence of rising cross-sections and the real decay rate of charged pions. An asymptotic ($x=\infty$) estimation of $E_{e\gamma}$ with the CORSIKA/QGSJET code at $E_0=10^{18}$ eV~\cite{Song} is shown in the figure as well.

%%%%%%%%%%%%%%%%%%%%%%%%%%%%%%%%%%%%%%%%%%%%%%%%%%%%%%%%%%%%%%%%%%%%%

\section{Conversion of the measured cosmic ray intensity to the spectrum of the primary beam}
\label{app-b}
There is a correction to be applied to the measured intensity of CRs before any comparison of the energy spectra observed by different EAS arrays.

The quantity $\hat{E}$ = 'primary particle energy' that has been estimated after a shower detection, and the actual energy of the particle, $E$, which has initiated the EAS, are different values, connected with each other by a relation to be found. Estimated energy has distribution around the given mean value $g(\hat{E},E)$, formed by the instrumental errors and fluctuations of the shower parameters with a RMS deviation, $\sigma$. Energy fluctuation is small in comparison with instrumental errors. Our aim here is to calculate the exact difference between the observed intensity of cosmic rays, $J(\hat{E})d\hat{E}$, and the original one $J(E)dE$ in the case of a rapidly falling power law spectrum.

The problem was solved, in general, by Zatsepin and Kalmykov in the previous century. The measured number of EAS particles, the so-called shower size, $N_e$, is connected with the primary energy. The function $g(N_e,E)$ depending on $E/N_e$ has been found by Zatsepin~\cite{Z}. Then Kalmykov has calculated the measured intensity in the case of a lognormal distribution of $N_e$~\cite{NNK}:

\begin{equation}
J(N_e)=J_0(N_e)exp(\frac{\sigma_N^2\kappa(\kappa-a_N)}{2a_N^2}),
\label{Eq:NNK}
\end{equation}
where $\sigma_N$ is RMS deviation of $\ln N_e$; $\kappa$ is spectrum index; $a_N=\frac{E}{N_e}\frac{dN_e}{dE}$.

Here we also assume the lognormal distribution of $y=\ln\hat{E}$  with the average value equal to $\ln E$. The observed intensity of cosmic rays is then given by the convolution of the primary spectrum, $J(z)=J_0\exp(-\kappa z)$, and the distribution of instrumental errors and fluctuations
\begin{equation}
\hat{J}(z)=\int_{-\infty}^\infty J(z-y)g(y)dy
=J_0\int_{-\infty}^\infty exp(-\kappa z+\kappa y)\frac{exp(-\frac{y^2}{2\sigma^2})}{\sqrt{2\pi}\sigma}dy.
\end{equation}
The resultant initial-to-observed intensity conversion factor is
$$R_J=\hat{J}(z)/J(z)=exp(\frac{\sigma^2\kappa^2}{2}).$$
The necessary conditions are a constant index and RMS error, or at least both changing only slowly with energy.

A distinct feature of the break in the spectrum is its shift along the energy scale. Due to the smeared transition between the conversion factors below and above the break, its position moves upward in energy.

\begin{figure}[t]
\center{\includegraphics[width=0.6\columnwidth]{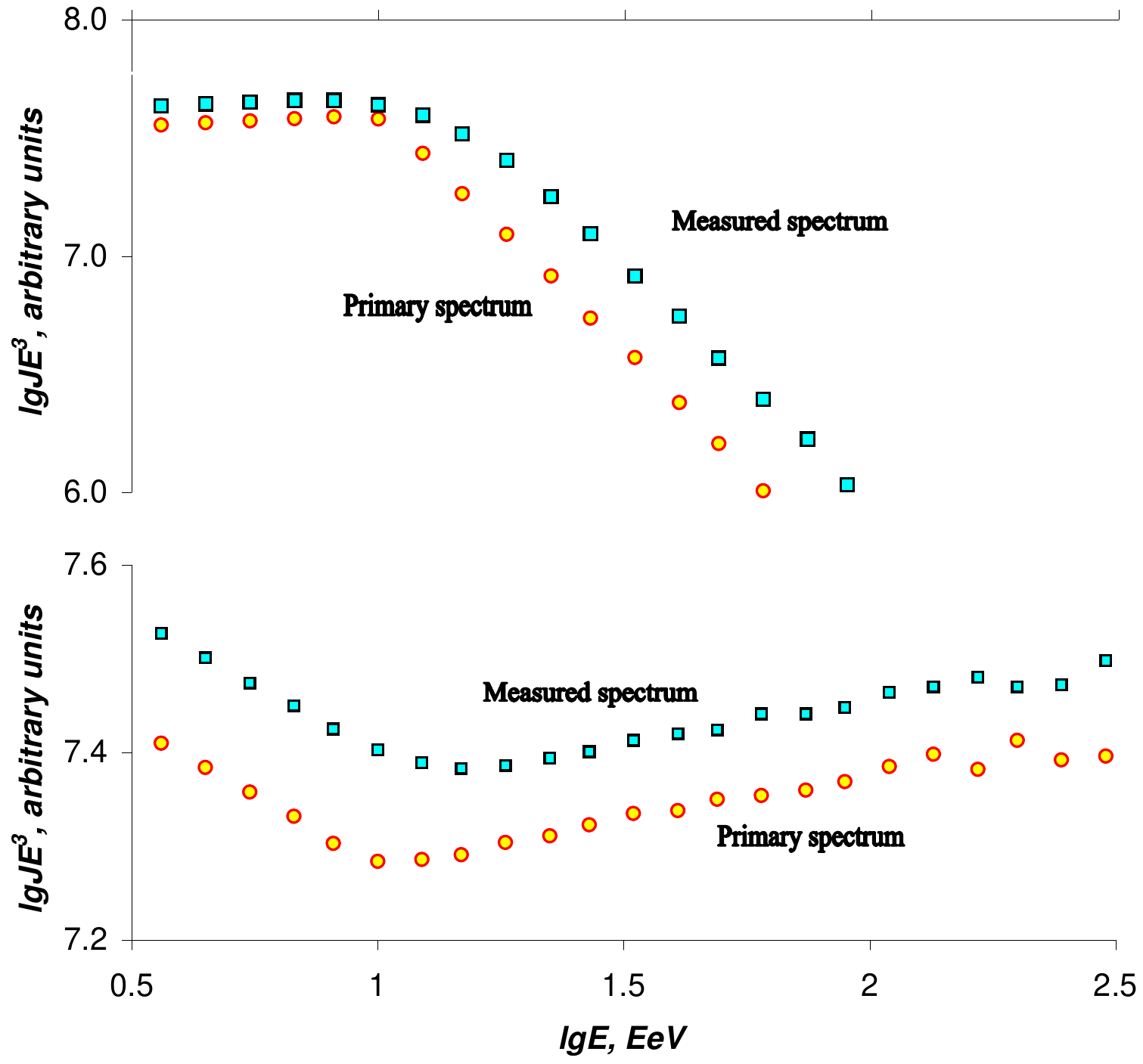}}
\caption{Simulation results of the spectrum measurement with ankle and knee features.}
\label{Fig:Breaks}
\end{figure}

To illustrate this we have simulated the primary energy spectrum with ankle and knee: random primary energies have been generated according to a broken power law with the index $\gamma=3.3$ and 2.9 below and above $10^{18}$ eV in the case of ankle, and $\gamma=2.9$ and 5 for the knee. The energy estimation is modelled by adding randomly a Gaussian error to $\ln E$ with $\sigma=0.32$. Figure~\ref{Fig:Breaks} shows the 'measured' and primary spectra. The different shifts in the logarithm of intensity on both sides of the break are clearly seen, as well as that the position of the observed break moves to the right.

%%%%%%%%%%%%%%%%%%%%%%%%%%%%%%%%%%%%%%%%%%%%%%%%%%%%%%%%%%%%%%%%%%%%%

\section{Data tables}
\label{app-c}
\begin{table}[h]
\begin{center}
\caption{Cosmic ray flux measured with the $C_2$ sub-array detectors.
Statistical errors are given where appreciable.}
\begin{tabular}{rr}
\hline
Energy, eV  & $J(E)$, m$^{-2}$\,sr$^{-1}$\,s$^{-1}$\,eV$^{-1}$ \\ \hline
 $1.60\times10^{15}$   & $8.55\times10^{-22}$ \\
 $2.00\times10^{15}$   & $4.63\times10^{-22}$ \\
 $2.50\times10^{15}$   & $2.56\times10^{-22}$ \\
 $2.90\times10^{15}$   & $1.80\times10^{-22}$ \\
 $3.50\times10^{15}$   & $1.05\times10^{-22}$ \\
 $4.60\times10^{15}$   & $4.83\times10^{-23}$ \\
 $5.70\times10^{15}$   & $2.32\times10^{-23}$ \\
 $6.30\times10^{15}$   & $1.68\times10^{-23}$ \\
 $7.80\times10^{15}$   & $9.06\times10^{-24}$ \\
 $9.50\times10^{15}$   & $5.25\times10^{-24}$ \\
 $1.15\times10^{16}$   & $2.83\times10^{-24}$ \\
 $1.40\times10^{16}$   & $1.60\times10^{-24}$ \\
 $1.78\times10^{16}$   & $7.27\times10^{-25}$ \\
 $2.19\times10^{16}$   & $3.33\times10^{-25}$ \\
 $2.80\times10^{16}$   & $1.73\times10^{-25}$ \\
 $3.50\times10^{16}$   & $(8.77\pm0.42)\times10^{-26}$ \\
 $4.20\times10^{16}$   & $(4.95\pm0.23)\times10^{-26}$ \\
 $5.10\times10^{16}$   & $(2.71\pm0.14)\times10^{-26}$ \\
 $6.60\times10^{16}$   & $(1.20\pm0.07)\times10^{-26}$ \\
 $7.90\times10^{16}$   & $(7.10\pm0.43)\times10^{-27}$ \\
 $1.00\times10^{17}$   & $(3.80\pm0.25)\times10^{-27}$ \\
 $1.66\times10^{17}$   & $(7.43\pm0.66)\times10^{-28}$ \\
 $2.00\times10^{17}$   & $(4.25\pm0.63)\times10^{-28}$ \\
 $3.10\times10^{17}$   & $(1.10\pm0.24)\times10^{-28}$ \\
\hline
\end{tabular}\label{Table:AutoData}
\end{center}\end{table}

\begin{table}[h]
\begin{center}
\caption{Cosmic ray flux measured with the $C_1$ sub-array detectors.}
\begin{tabular}{rr}
\hline
Energy, EeV  & $J(E)$, m$^{-2}$\,sr$^{-1}$\,s$^{-1}$\,eV$^{-1}$ \\ \hline
 $2.60\times10^{17}$   & $2.16\times10^{-28}$ \\
 $4.10\times10^{17}$   & $5.91\times10^{-29}$ \\
 $6.50\times10^{17}$   & $1.60\times10^{-29}$ \\
 $1.00\times10^{18}$   & $4.20\times10^{-30}$ \\
 $1.27\times10^{18}$   & $(1.35\pm0.07)\times10^{-30}$ \\
 $2.02\times10^{18}$   & $(3.34\pm0.23)\times10^{-31}$ \\
 $3.22\times10^{18}$   & $(7.95\pm0.87)\times10^{-32}$ \\
 $5.08\times10^{18}$   & $(1.69\pm0.37)\times10^{-32}$ \\
 $8.06\times10^{18}$   & $(3.35\pm1.07)\times10^{-33}$ \\
 $1.27\times10^{19}$   & $(1.52\pm0.75)\times10^{-33}$ \\
 $2.02\times10^{19}$   & $(4.39\pm2.54)\times10^{-34}$ \\
 $3.22\times10^{19}$   & $(1.69\pm0.99)\times10^{-34}$ \\
 $5.08\times10^{19}$   & $(1.41\pm1.01)\times10^{-34}$ \\
\hline
\end{tabular}\label{Table:MainData}
\end{center}\end{table}

%%%%%%%%%%%%%%%%%%%%%%%%%%%%%%%%%%%%%%%%%%%%%%%%%%%%%%%%%%%%%%%%%%%%%


\section*{References}
\begin{thebibliography}{99}
\bibitem{Jelley} Galbraith W and Jelley J 1953 {\it Nature} {\bf 171} 349
\bibitem{Chudakov} Nesterova N M and Chudakov A E 1955 {\em JETP} {\bf 28} 384
\bibitem{Mono} Dyakonov M N et al. 1991 {\it Cosmic Rays of Extremely High Energy} (Novosibirsk: Nauka)
\bibitem{CERN} Ivanov A A, Knurenko S P and Sleptsov I.Ye 2003 {\it Nucl.Phys. B (Proc. Suppl.)} {\bf 122} 226
\bibitem{Autonom} Afanasiev B N et al. 1996 {\it Proc. ISEHECR: Astrop. and Future Observ. (Tokyo)} 412
\bibitem{Sleptsov} Sleptsov I Ye 1974 Thesis (Lebedev Physical Institute, Moscow)
\bibitem{Hara} Hara T et al. 1977 {\it Proc. 15$^{th}$ Int. Cosmic Ray Conf. (Plovdiv)} {\bf 8} 308
\bibitem{Nesterova} Nesterova N M 1961 Thesis (Moscow State University, Moscow)
\bibitem{Glushkov} Glushkov A V 1982 Thesis (Moscow State University, Moscow)
\bibitem{Tamm} Frank I M and Tamm I E 1937 {\it C. R. Acad. Sci. URSS} {\bf 14} 109
\bibitem{LaJolla} Dyakonov M N et al. 1985 {\it Proc. 12$^{th}$ Int. Cosmic Ray Conf. (La Jolla)} {\bf 2} 194
\bibitem{Pampa} Mostafa M A et al. 2003 {\it Proc. 28$^{th}$ Int. Cosmic Ray Conf. (Tsukuba)} {\bf 1} 465
\bibitem{DensFluct} Dyakonov M N et al. 1975 {\it Proc. 14$^{th}$ Int. Cosmic Ray Conf. (Munchen)} {\bf 12} 4339
\bibitem{HPCher} Hammond R T et al. 1977 {\it Proc. 15$^{th}$ Int. Cosmic Ray Conf. (Plovdiv)} {\bf 8} 281
\bibitem{Dyakonov} Dyakonov M N 1981 Thesis (Institute for Nuclear Research, Moscow)
\bibitem{Ivanenko} Ivanenko I P et al. 1979 {\it Proc. 16$^{th}$ Int. Cosmic Ray Conf. (Kyoto)} {\bf 9} 83
\bibitem{LagutinLDF} Lagutin A A et al. 1987 {\it Preprint} No 1289 (Leningrad: Konstantinov Institute for Nuclear Research)
\bibitem{Ddnk} Dedenko L G et al. 2005 {\it Proc. 29$^{th}$ Int. Cosmic Ray Conf. (Pune)} {\bf 7} 219
\bibitem{JETP07} Ivanov A A, Knurenko S P and Sleptsov I Ye 2007 {\it JETP} {\bf 104} 872
\bibitem{Prav04} Pravdin M I et al. 2004 {\it Izv. Akad. Nauk. Ser. Fiz.} {\bf 68} 1621
\bibitem{Khrenov} Khrenov B A 1986 Thesis (Moscow State University, Moscow)
\bibitem{Eem} Knurenko S P, Ivanov A A, Sleptsov I E and Sabourov A V 2006 {\it JETP Lett.} {\bf 83} 473
\bibitem{Song} Song C et al. 2000 {\it Astropart. Phys.} {\bf 14} 7
\bibitem{Iv01} Ivanov A A et al. 2001 {\it Izv. Akad. Nauk. Ser. Fiz.} {\bf 65} 1221
\bibitem{Blanca} Fowler J W et al. 2001 {\it Astropart. Phys.} {\bf 15} 49
\bibitem{Tunka} Chernov D et al. 2004 {\it Int. J. Mod. Phys. A} {\bf 20} 6799
\bibitem{Nagano} Nagano M and Watson A A 2000 {\it Rev. Mod. Phys.} {\bf 72} 689
\bibitem{Khristi} Khristiansen G B and Kulikov G V 1958 {\it JETP} {\bf 35} 635
\bibitem{Akeno} Nagano M et al. 1992 {\it J. Phys. G: Nucl. Phys.} {\bf 18} 423
\bibitem{Kascade} Antoni T et al. 2005 {\it Astropart. Phys.} {\bf 24} 1
\bibitem{Tibet} Amenomori M et al. 2008 {\it Astrophys. Journ.} submitted; {\it Preprint} arXiv:hep-ex/0801.1803
\bibitem{Berezhko} Berezhko E G and V\"olk H J 2007 {\it Preprint} arXiv:astro-ph/0704.1715
\bibitem{Kalmykov} Kalmykov N N and Pavlov A I 1999 {\it Proc. 26$^{th}$ Int. Cosmic Ray Conf. (Salt Lake City)} {\bf 4} 263
\bibitem{SSM1} Erlykin A D and Wolfendale A W 2001 {\it J. Phys. G: Nucl. Part. Phys.} {\bf 27} 1005
\bibitem{Lagutin} Lagutin A A, Nikulin Yu A and Uchaikin V V 2001 {\it Nucl. Phys. B (Proc. Suppl.)} {\bf 97} 267
\bibitem{SSM2} Erlykin A D 2005 {\it Int. J. Mod. Phys. A} {\bf 20} 6584
\bibitem{LagutinSSM} Lagutin A A et al. 2007 {\it Izv. Akad. Nauk. Ser. Fiz.} {\bf 71} 605
\bibitem{CRIS} Egorova V P et al. 2004 {\it Nucl. Phys. B (Proc. Suppl.)} {\bf 136} 3
\bibitem{AGASASpectrum} Takeda M et al. 2003 {\it Astropart. Phys.} {\bf 19} 447
\bibitem{HiRes08} Abbasi R U 2008 {\it Phys. Rev. Lett.} 100:101101
\bibitem{PAO08} PAO Collaboration 2008 {\it Phys. Rev. Lett.} 101:061101
\bibitem{Carvalho} Carvalho W Jr, Albuquerque I F M and de Souza V 2007 {\it Astropart. Phys.} {\bf 28} 89
\bibitem{Merida} Ivanov A A, Knurenko S P and Sleptsov I Ye 2007 {\it Proc. 30$^{th}$ Intern. Cosmic Ray Conference (Merida)}
\bibitem{Bahcall} Bahcall J N and Waxman E 2003 {\it Phys. Lett.} {\bf B556} 1
\bibitem{Dip07}Berezinsky V, Gazizov A Z and Grigorieva S I 2006 {\it Phys. Rev. D} {\bf 74} 043005
\bibitem{MS} De Marco D and Stanev T 2005 {\it Phys. Rev. D} {\bf 72} 081301
\bibitem{WgW} Wibig T and Wolfendale A W 2005 {\it J. Phys. G: Nucl. Phys.} {\bf 31} 255
\bibitem{Z} Zatsepin G T 1959 {\it Proc. Intern. Cosmic Ray Conference (Moscow)} {\bf 2} 212
\bibitem{NNK} Kalmykov N N 1969 {\it Yad. Fiz.} {\bf 10} 121

\end{thebibliography}
\end{document}